\documentclass[10pt,conference]{IEEEtran}
\usepackage{amsmath,amssymb,amsfonts}
\usepackage{algorithmic}
\usepackage{graphicx}
\usepackage{textcomp}
\usepackage{xcolor}
\usepackage[hyphens]{url}
\usepackage{fancyhdr}

\newcommand{\hpcayear}{2026}

\newcommand{\hpcasubmissionnumber}{186}

\def\hpcacameraready{} %

\newcommand\hpcaauthors{Hongshi Tan$^\dagger$, Yao Chen$^\dagger$, Xinyu Chen$^\ddagger$, Qizhen Zhang$^\S$, Cheng Chen$^{\dagger\dagger}$, Weng-Fai Wong$^\dagger$, Bingsheng He$^\dagger$}
\newcommand\hpcaaffiliation{National University of Singapore$^\dagger$,
The Hong Kong University of Science and Technology (Guangzhou)$^\ddagger$,
\\ University of Toronto$^\S$, ByteDance$^{\dagger\dagger}$}
\newcommand\hpcaemail{hongshi@u.nus.edu, yaochen@nus.edu.sg, xinyuchen@hkust-gz.edu.cn, qz@cs.toronto.edu, \\ chencheng.sg@bytedance.com, wongwf@comp.nus.edu.sg, hebs@comp.nus.edu.sg}

\usepackage{amsfonts}
\usepackage{colortbl}
\usepackage{xspace}
\usepackage{lipsum}

\usepackage{cite}

\usepackage{fontawesome}

\usepackage{caption}
\usepackage{subcaption}

\usepackage[]{hyperref}

\usepackage{enumitem}

\usepackage[export]{adjustbox}

\usepackage{lineno}

\usepackage{xcolor}
\usepackage{soul}        %

\colorlet{pink}{red!40}
\colorlet{cyanblue}{cyan!80}

\definecolor{myred}{RGB}{250,214,221}
\definecolor{myblue}{RGB}{190,228,254}

\usepackage{tikz}
\usetikzlibrary{calc}

\usepackage{pgf}

\newdimen\mydim

\newcommand{\gettikzxy}[3]{%
  \tikz@scan@one@point\pgfutil@firstofone#1\relax
  \edef#2{\the\pgf@x}%
  \edef#3{\the\pgf@y}%
}

\newdimen\XCoord
\newdimen\YCoord

\newcommand*{\tikzmk}[1]{\tikz[remember picture,overlay,blend mode=multiply] \node (#1) {};\ignorespaces}

\usepackage{mathtools}
\usepackage{bm}
\usepackage{empheq}
\usepackage{amsthm}

\usepackage{resizegather} %

\usepackage[linguistics]{forest}

\usepackage[short]{optidef}

\newtheorem{theorem}{Theorem}[section]

\usepackage{siunitx}
\newcommand{\unsim}{\mathord{\sim}}

\usepackage[ruled,vlined,linesnumbered,resetcount,algosection,nofillcomment]{algorithm2e}

\SetKwInput{KwData}{Input}
\SetKwInput{KwResult}{Output}
\SetKwFor{Pfor}{parallel for}{}{}
\SetKwFor{PipFor}{pipelined for}{}{}
\SetKwFor{PipForEach}{pipelined foreach}{do}{endfch}
\SetKw{KwBy}{by}
\SetKw{KwNot}{not}
\SetKw{KwBreak}{break}
\SetKwFor{Loop}{loop}{}{}

\SetKwRepeat{Do}{do}{while}
\SetKwRepeat{PipDo}{pipelined do}{while}
\SetKwProg{Pn}{function}{:}{\KwRet}
\SetKw{Output}{output}

\let\oldnl\nl%
\newcommand{\nonl}{\renewcommand{\nl}{\let\nl\oldnl}}%

\definecolor{mycommentcolor}{rgb}{0.0, 0.5, 0.6}

\SetCommentSty{mycommfont}

\usepackage{comment}

\usepackage[normalem]{ulem}

\newcommand{\myrevision}{false}
\newcommand{\myrevisionbigpage}{false}

\usepackage{ifthen}

\ifthenelse{\equal{\myrevisionbigpage}{true}}
{
  \paperwidth=\dimexpr \paperwidth + 8cm\relax
  \oddsidemargin=\dimexpr\oddsidemargin + 4cm\relax
  \evensidemargin=\dimexpr\evensidemargin + 4cm\relax
  \marginparwidth=\dimexpr \marginparwidth + 4cm\relax

  \paperheight=\dimexpr \paperheight + 8cm\relax
}
{
}

\ifthenelse{\equal{\myrevision}{true}}
  {%
\usepackage[colorinlistoftodos]{todonotes}
\setuptodonotes{fancyline, color=blue!30, size=\small}

    \newcommand\td[1]{\todo{#1}}
    \newcommand\tds[1]{\todo{\sout{#1}}}
    \newcommand\suggestion[2]{\todo[backgroundcolor=green!20!white]{ #2}}
    \newcommand\solvedsuggestion[2]{\todo[backgroundcolor=green!20!white]{\sout{#2}}}

    \newcommand\rvh[2]{\xspace{\color{blue}%
\saveacmlinepos{acmlinenostart:#1}\label{acmline:#1}#2\saveacmlinepos{acmlinenoend:#1}}\xspace}

  }
  {%
    \newcommand\td[1]{}
    \newcommand\tds[1]{}
    \newcommand\suggestion[2]{}
    \newcommand\solvedsuggestion[2]{}

    \newcommand\rvh[2]{{#2}}
  }

\newcommand{\setword}[2]{%
  \phantomsection
  #1\def\@currentlabel{\unexpanded{#1}}\label{#2}%
}

\usepackage{etoolbox}  %

\usepackage[savepos,totpages,titleref,dotfill,counter,user]{zref}
\usepackage{pageslts}
\usepackage{atbegshi}     %
\usepackage{xfp}  %

\makeatletter

\newcommand{\refacmlineno}[1]{%
  \edef\Ysp{\zposy{acmlinenostart:#1}}%
  \edef\Xsp{\zposx{acmlinenostart:#1}}%
  \edef\Psp{\zref@extract{zlabel:acmlinenostart:#1}{abspage}}%
  \edef\Yep{\zposy{acmlinenoend:#1}}%
  \edef\Xep{\zposx{acmlinenoend:#1}}%
  \edef\Pep{\zref@extract{zlabel:acmlinenoend:#1}{abspage}}%
  \hyperref[acmline:#1]{
  {\textcolor{red}{Line $\fpeval{ (\Psp - 2) * 58 * 2 + (round((\Xsp + 10457825) / 20915650) - 1) * 58  +  round((580 + 121 - round((\Ysp / 65536) )) /11) + 1  }\unsim \fpeval{ (\Pep - 2) * 58 * 2 + (round((\Xep + 10457825) / 20915650) - 1) * 58 +  round((580 + 121 - round((\Yep / 65536) )) /11) + 1}$}%
  }%
  }%
}

\newcommand{\refsection}[1]{\hyperref[acmline:#1]{\textcolor{red}{Section \zref{zlabel:acmlinenostart:#1}}}}

\makeatother

\usepackage{booktabs}
\usepackage{tabularx}
\usepackage{multirow,multicol,threeparttable, tablefootnote}
\usepackage{makecell}
\newcolumntype{Y}{>{\centering\arraybackslash}X}

\usepackage{mdframed}
\mdfdefinestyle{myframe}{%
    skipabove=0pt,
    skipbelow=0pt
}

\usepackage[most]{tcolorbox}

\usepackage{cleveref}
\newcommand{\mref}[2]{\textcolor{red}{\hyperref[#1]{#2}}}
\crefname{algocf}{alg.}{algs.}
\Crefname{algocf}{Algorithm}{Algorithms}

\usepackage{pifont}
\def\circled#1{%
        \ding{\numexpr201+#1\relax}%
}

\pgfmathsetmacro{\nodebasesize}{1} %
\pgfmathsetmacro{\nodeinnersep}{0.05}

\newcommand{\mysys}{RidgeWalker\xspace}

\newcommand{\graphG}{\mathcal{G}}

\newcommand{\vrfigsize}{-12pt}

\title{\mysys{}: Perfectly Pipelined Graph Random Walks on FPGAs}

\author{
  \ifdefined\hpcacameraready
    \IEEEauthorblockN{\hpcaauthors{}}
      \IEEEauthorblockA{
        \hpcaaffiliation{} \\
        \hpcaemail{}
      }
  \else
    \IEEEauthorblockN{\normalsize{HPCA \hpcayear{} Submission
      \textbf{\#\hpcasubmissionnumber{}}} \\
      \IEEEauthorblockA{
        Confidential Draft \\
        Do NOT Distribute!!
      }
    }
  \fi 
}

\fancypagestyle{camerareadyfirstpage}{%
  \fancyhead{}
  \fancyfoot[C]{}
}
\begin{document}
\maketitle

\ifdefined\hpcacameraready
  \thispagestyle{camerareadyfirstpage}
  \pagestyle{empty}
\else
  \thispagestyle{plain}
  \pagestyle{plain}
\fi

\newcommand{\hpcaheight}{0mm}
\ifdefined\eaopen
\renewcommand{\hpcaheight}{12mm}
\fi

\begin{abstract}

Graph Random Walks (GRWs) offer efficient approximations of key graph properties and have been widely adopted in many applications.
However, GRW workloads are notoriously difficult to accelerate due to their strong data dependencies, irregular memory access patterns, and imbalanced execution behavior. While recent work explores FPGA-based accelerators for GRWs, existing solutions fall far short of hardware potential due to inefficient pipelining and static scheduling.
This paper presents \mysys{}, a high-performance GRW accelerator designed for datacenter FPGAs. The key insight behind \mysys{} is that the Markov property of GRWs allows decomposition into stateless, fine-grained tasks that can be executed out-of-order without compromising correctness. Building on this insight, \mysys{} introduces an asynchronous pipeline architecture with a feedback-driven scheduler grounded in queuing theory. This design enables perfect pipelining and adaptive load balancing. 
We prototype \mysys{} on FPGAs and evaluate its performance across a range of GRW algorithms and real-world graph datasets. Experimental results demonstrate that \mysys{} achieves an average speedup of $7.0\times$ over state-of-the-art FPGA solutions and $8.1\times$ over GPU solutions, with peak speedups of up to $71.0\times$ and $22.9\times$, respectively. The source code is publicly available at \underline{\textit{\url{https://github.com/Xtra-Computing/RidgeWalker}}}.

\end{abstract}

\section{Introduction}
\label{sec:introduction}

\begin{figure}[t]
\centering
\includegraphics[width=\columnwidth]{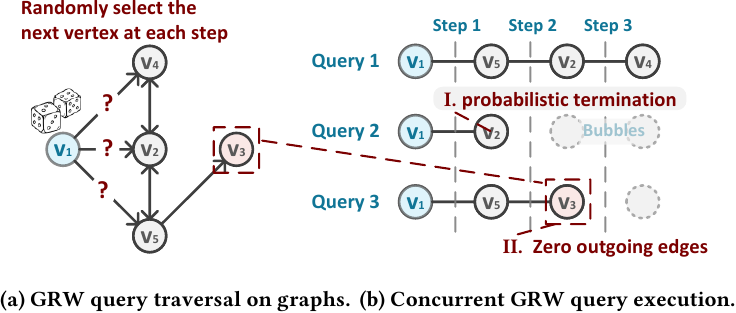}
\caption{Illustration of the inherent randomness and imbalanced traversal in GRWs.}
\label{fig:async_exec}
\end{figure}

Modern data science and network analysis applications increasingly rely on {\em graph random walks} (GRWs) to efficiently explore and characterize complex network structures. GRW algorithms, based on Markov processes~\cite{lambiotte2015random}, traverse a graph by probabilistically selecting neighbors based solely on the current vertex, as shown in~\Cref{fig:async_exec}a. This sampling strategy approximates key graph characteristics, such as centrality, influence propagation, and community structure, without exhaustive traversal~\cite{10.1145/3318464.3380562, liao2023efficient, yu2014personalized, perozzi2014deepwalk}. 
Therefore, GRWs have been used as the core building blocks in databases~\cite{miller2013graph,10.1145/3543507.3583474}, machine learning~\cite{dong2017metapath2vec, grover2016node2vec, cappelletti2023grape, dgl}, and graph-based reasoning~\cite{liu2016hierarchical, wang2020learning,jimenez2024hipporag}.

GRWs have become the dominant performance bottleneck in many graph-learning workloads, often accounting for over 50\% of total execution time~\cite{gong2023gsampler}.
This has spurred extensive research efforts aimed at developing high-performance GRW solutions on CPUs and GPUs~\cite{sun2021thunderrw, yang2019knightking, gong2023gsampler, 10.1145/3588944, yin2022scalable, jangda2021accelerating}.
However, general-purpose architectures like GPUs, designed for workloads with massive regular parallelism, are ill-suited for GRWs, which feature irregular computation and highly random memory access patterns~\cite{wang2023optimizing}. 
As a result, even state-of-the-art GPU-based GRW~\cite{gong2023gsampler} utilizes only 0.9\% of total random access memory bandwidth (see \Cref{sub:comparison_gpu}), exposing a fundamental mismatch between GRW workloads and conventional hardware.

This mismatch highlights the need for domain-specific architectures tailored to the unique characteristics of GRWs for improved bandwidth efficiency and performance. 
With architectural reconfigurability and access to {\em high-bandwidth memory} (HBM), modern FPGAs have attracted growing interest~\cite{gao2023fastrw,10.1145/3588944,su2021graph} for building customized GRW accelerators and adapting to rapidly evolving GRW applications.
However, despite this potential, existing FPGA-based GRW accelerators still fall short of fully utilizing the hardware’s capabilities.  
This performance gap stems from two inherent characteristics in GRW workloads that make high-performance accelerator design particularly difficult.

First, \textbf{GRWs exhibit strong data dependencies and unpredictable random memory accesses.} 
Each step of a walk requires a random access to retrieve the neighbors of the current vertex, and the next step cannot proceed until this retrieval is complete (shown in~\Cref{fig:async_exec}a). 
Like traditional graph processing accelerators~\cite{chen2021thundergp,chen2022regraph}, existing GRW design~\cite{gao2023fastrw} caches frequently accessed vertices on-chip based on their access frequency.
However, such strategies are largely ineffective for GRWs due to their probabilistic neighbor selection. As illustrated in~\Cref{fig:async_exec}a, even walks starting from the same vertex \(v_1\) rapidly diverge along distinct paths, offering minimal locality and thus limited cache reuse. Our profiling of FastRW~\cite{gao2023fastrw} shows that once the graph exceeds on-chip cache capacity, random access memory bandwidth utilization drops by up to 42\%.

Second, \textbf{concurrent GRW query execution introduces significant workload imbalance.} 
The lengths of each query vary greatly due to probabilistic termination and the structural sparsity of real-world graphs. For instance, vertex \(v_3\) in~\Cref{fig:async_exec}b  has no outgoing edges, forcing any walk that reaches it to terminate immediately. Such behavior leads to unpredictable execution times across queries. 
Existing GRW architectures~\cite{gao2023fastrw,10.1145/3588944,su2021graph} rely on static scheduling and fail to adapt to runtime imbalances,  leading to pipeline bubbles and under-utilization of the compute pipeline. 
Our evaluation shows that under such an imbalance, the effective random access memory bandwidth utilization can drop below 2.3\% of the hardware’s theoretical peak (see~\Cref{sub:inefficiency}).

We attribute the performance gap of existing designs to the lack of \emph{perfectly pipelined} execution~\cite{stasytis2023optimization}, which requires (a) proactively resolving data dependencies and (b) eliminating pipeline bubbles.
Our key insight is that the Markovian nature of GRWs confines inter-step data dependencies to the current vertex, allowing queries to be decomposed into stateless, minimal tasks for fine-grained execution without compromising statistical correctness. 
This property motivates us to design a GRW accelerator architecture that supports out-of-order, asynchronous query execution to hide memory latency and adapts dynamically to workload imbalance. 

In this paper, we present \mysys{}, an efficient FPGA-based GRW accelerator that leverages the Markov property of GRWs to enable fine-grained task decomposition and achieve perfectly pipelined execution. 
First, \mysys{} enables out-of-order GRW query execution through an asynchronous pipeline architecture. By interleaving tasks from different queries and executing them independently, queries are dynamically scheduled and redistributed across pipelines at runtime rather than being constrained by sequential input order or affinity to the processing pipeline. This flexible execution model effectively hides the latency of random memory accesses.
Second, \mysys{} incorporates a feedback-driven scheduler that adaptively resolves workload imbalance and pipeline bubbles.
Grounded in {\em queuing theory}, it provides formal guarantees of zero pipeline bubbles, sustaining continuous data flow even under highly imbalanced GRW workloads.
To summarize, this paper makes the following contributions:

\begin{itemize}[leftmargin=*]
\setlength\itemsep{-0.5mm}
  \item We propose \mysys{}, the first GRW accelerator to achieve perfectly pipelined execution, aligning architectural design with the Markovian structure of GRW algorithms.

 \item We introduce an asynchronous GRW accelerator architecture that efficiently pipelines decomposed stateless GRW tasks, achieving fine-grained concurrency and near-optimal random-access bandwidth utilization.

  \item We introduce a feedback-driven scheduler based on queuing theory, offering formal guarantees for zero pipeline bubbles and sustained high utilization under irregular workloads and high parallelism.

  \item We implement \mysys{} on various FPGAs and demonstrate its generality and performance across multiple GRW algorithms and datasets, achieving up to $71.0\times$ and $22.9\times$ speedup over prior FPGA and GPU solutions, respectively.
\end{itemize}

\section{Background}

\subsection{CSR Graph Representation}

\begin{figure}[t!]
\centering
\includegraphics[width=\linewidth]{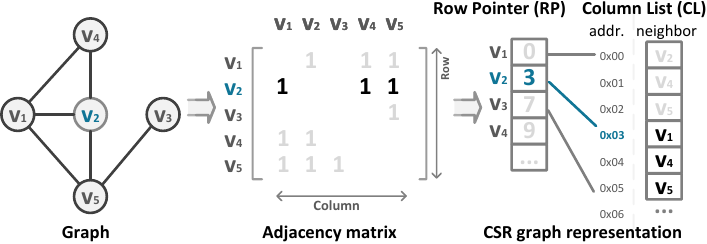}
\caption{An example of graph representation in CSR format.}
\label{fig:csr}
\end{figure}

\Cref{fig:csr} illustrates an example graph encoded in the \emph{compressed sparse row} (CSR) format, the adjacency representation most commonly used in GRW workloads. CSR uses two arrays: the row pointer array (RP) and the column list array (CL). Each entry $\text{RP}[i]$ specifies the starting offset of vertex $v_i$’s neighbor list in CL, which stores the column indices of non-zero entries in the adjacency matrix. For example, $\text{RP}[2] = 3$ indicates that the neighbors of $v_2$ begin at CL address \texttt{0x03}. Storing every vertex’s neighbors contiguously allows an \(O(1)\) index lookup, making it well-suited and widely adopted for the random sampling in GRWs~\cite{sun2021thunderrw, yang2019knightking, gong2023gsampler, 10.1145/3588944, yin2022scalable, jangda2021accelerating}.

\begin{algorithm}[tb]
\SetInd{0.1em}{2em}
\KwData{$\graphG$: a given graph,
$\mathbb{Q}$: a set of input queries.
}
\KwResult{$\textit{res}$: paths of traversed vertices.}
\tikzmk{Zero}
$\textit{res} = \emptyset$\;
\ForEach{$Q\in \mathbb{Q}$}{
    $v_{\textit{curr}}= Q.v_{\textit{start}}$,
    $\textit{path} = \emptyset$\;
\Loop{}{
    \tcc{Access the row pointer of CSR;}
    $\{\textit{addr}, \textit{deg}\} = \text{\textbf{row\_access}}(v_\textit{curr},\graphG)$\;\label{line:1_rpa}
    \tcc{Application-specific sampling;}
    $\textit{index} = \text{\textbf{sampling}}(\textit{addr}, \textit{deg},\graphG)$\;\label{line:1_sample}
    \tcc{Access the column list of CSR;}
    $v_\textit{curr} = \text{\textbf{\textbf{column\_access}}}(\textit{addr}, \textit{index},\graphG)$\;\label{line:1_cla}
    $\textit{path}.\textbf{push}(v_\textit{curr})$\; \label{line:1_update}
    \If{$(Q.{\normalfont \textbf{is\_end}}())$}
    {
        $\textit{res}.\textbf{push}(\textit{path})$\;\label{line:1_write}
        \KwBreak \;
    }
}
}
\Return{$\textit{res}$}\;
\caption{General GRW algorithm}\label{alg:grw}
\end{algorithm}

\subsection{GRW Execution Flow}
\Cref{alg:grw} outlines the execution flow of a general GRW algorithm.
The input is a CSR graph \(\graphG\), and the application launches a set of random walk queries \(\mathbb{Q}\) with designated starting vertices.
A query begins at \(v_{\textit{curr}} = Q.v_{\textit{start}}\) and advances one step at a time.
For each step, the algorithm first reads the degree of \(v_{\textit{curr}}\) and the pointer to its neighbor list from the row pointer array of the CSR graph (Line~\ref{line:1_rpa}).
An application-specific sampling function then chooses a neighbor index within that list (Line~\ref{line:1_sample}).
Using the selected index, the algorithm accesses the column list to fetch the next vertex to visit.
The walk terminates when it either reaches the maximum length or encounters a vertex with no outgoing edges.

GRWs are naturally difficult to perfectly pipeline, because their execution cannot be structured as a perfectly nested loop~\cite{dave2019dmazerunner}, since the number of inner loop iterations is not known in advance. This unpredictability prevents the straightforward adoption of conventional pipelining techniques that rely on static scheduling, such as Finite State Machine with Datapath (FSMD)~\cite{davis2022finite}, SDC-based modulo scheduling~\cite{6691121}, and similar approaches, from achieving perfect pipelining and optimal bandwidth utilization.

\section{Motivation and Design Principles}
\label{sub:inefficiency}

\subsection{The Need for Specialized Accelerators for GRWs}
\label{sub:the_need_for_specialized_accelerators_for_grws}

While extensive research has focused on accelerators for general graph processing~\cite{chen2021thundergp,jaiyeoba2023acts,jaiyeoba2024dynamic,jaiyeobaswift,zhou2019hitgraph,hu2021graphlily,GraphPulse2020,chen2022regraph,JetStream2021,ScalaGraph2022,chen2022thundergpr}, these designs are fundamentally mismatched with the needs of GRWs.
General graph processing typically involves scanning all neighbors of a vertex to perform aggregation or message propagation. This has motivated optimizations like update coalescing (e.g., ACTS~\cite{jaiyeoba2023acts}, Swift~\cite{jaiyeobaswift}) and sequential buffering (e.g., GraphPulse~\cite{GraphPulse2020}) that efficiently utilize on-chip memory. However, GRWs are structurally different, as they traverse the graph by sampling a single, randomly selected neighbor at each step, without aggregation. This sparsity in access renders such buffering and aggregation techniques ineffective.
As a result, there is a pressing need for specialized architectures that can match the stochastic, data-dependent behavior of GRWs.

\subsection{Inefficiencies in Existing GRW Accelerators}

\begin{figure}[t]
\centering
\includegraphics[width=1\columnwidth]{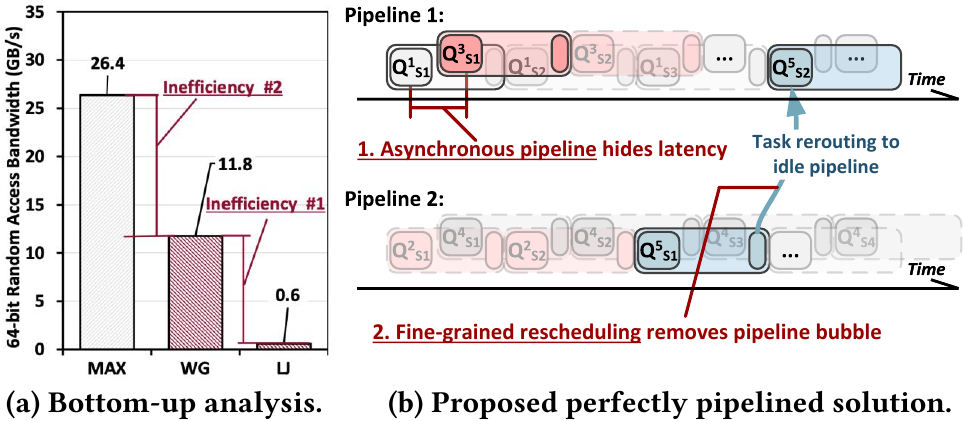}
\caption{
(a) Bandwidth analysis of SOTA FPGA accelerator FastRW~\cite{gao2023fastrw}, highlighting underutilization.
(b) Perfectly pipelined parallel GRW execution on two pipelines, {$Q_{sy}^x$ denotes the $y$-th step of traversal for query $x$.}}
\label{fig:limitations}
\end{figure}

While several prior works have explored FPGA-based GRW acceleration~\cite{gao2023fastrw,10.1145/3588944,su2021graph}, current designs fall far short of the hardware’s potential.
To explore design challenges of GRW accelerators, we conduct in-depth performance analysis for FastRW~\cite{gao2023fastrw} and LightRW~\cite{10.1145/3588944}.
We measure its bandwidth utilization as $\smash{B_{\text{measured}} / B_{\text{peak}}}$, where $B_{\text{measured}}$ is the effective bandwidth observed during GRW traversal and is calculated by dividing the total memory footprint of traversed edges by the overall execution time. $B_{\text{peak}}$ denotes the theoretical peak bandwidth provided by DRAM or HBM, estimated following Asifuzzaman~\textit{et al.}~\cite{asifuzzaman2021demystifying}. Since each GRW step typically triggers a DRAM row-buffer miss, we consider this to compute peak 64-bit random-access bandwidth $B_{\text{peak}}$, aligned with the granularity of vertex-level random accesses in GRWs:
\begin{equation}
B_\text{peak}=\frac{f_\text{mem}}{t_\text{RRD}} \times N_{\text{chn}} \times \frac{64\text{-bit}}{8},\label{eq:dram}
\end{equation}
where $f_\text{mem}$ is the memory operating frequency, $t_\text{RRD}$ is the row-to-row delay, and $N_{\text{chn}}$ denotes the number of available memory channels. We conclude two key observations:

\noindent\textbf{Observation \#1: Memory access latency is inherently difficult to hide due to GRW data dependency.}
\Cref{fig:limitations}a shows that FastRW~\cite{gao2023fastrw} sustains 11.8 GB/s of effective bandwidth on the small \texttt{WG} graph, where the row-pointer array fits entirely in on-chip SRAM. 
However, performance drops drastically to 0.6 GB/s, merely 2.3\% of peak, on the larger \texttt{LJ} graph.
This drop reveals a fundamental challenge: GRWs impose strict, step-by-step data dependencies, where each step must complete a random memory access to determine the next vertex before proceeding. These accesses are not only random but also sequentially chained, making it extremely difficult to prefetch or parallelize across steps. 
\textit{Thus, optimizing the memory hierarchy alone is insufficient; instead, resolving GRW’s inter-step dependencies is essential for hiding latency and achieving high performance.}

\noindent\textbf{Observation \#2: Static scheduling cannot help the workload imbalance problem in GRW.}
\Cref{fig:limitations}a compares FastRW’s observed bandwidth with the theoretical peak $\texttt{MAX}$ derived in~\Cref{eq:dram}. Even when the row-pointer array fits entirely within the on-chip RAM, FastRW reaches only 45\% of the peak. The shortfall is not a memory issue but query scheduling. Individual queries finish at different times, so idle cycles accumulate while the fixed schedule waits for the slowest query. 
LightRW~\cite{10.1145/3588944} improves locality by batching queries in a ring buffer, yet still issues every step in a predetermined order; when a walk terminates early, its reserved slots remain empty. 
Our analysis on LightRW reveals bubble ratios up to 37\%, confirming that static scheduling cannot cope with GRW’s highly imbalanced runtime behavior. 
\textit{To close this gap, a scheduler must adapt on the fly, redirecting ready tasks to a free processing pipeline, to keep all hardware resources busy.
}

\begin{figure*}[t]
\centering
\includegraphics[width=\linewidth]{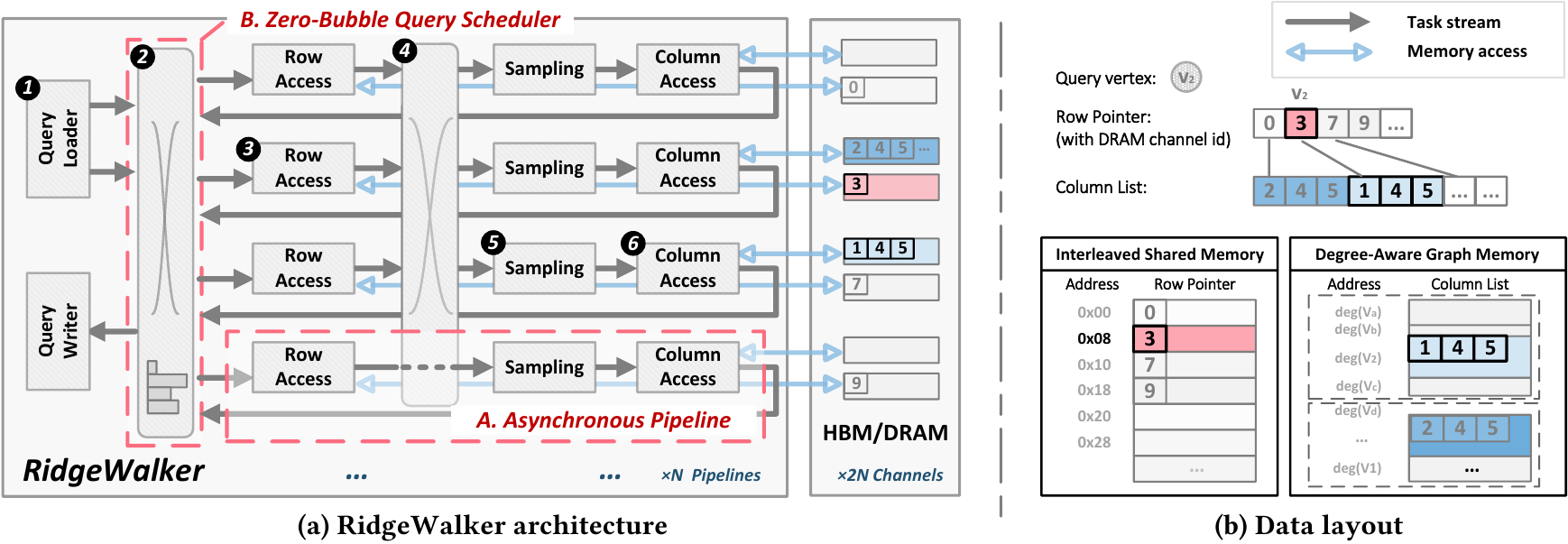}
\caption{\mysys~architecture overview.}
\label{fig:architecture_overview}
\end{figure*}

\subsection{\mysys{}: Towards Perfectly Pipelined GRWs}
\label{sub:_mysys_}
Our analysis shows that efficient GRW acceleration requires rethinking both the algorithm and architecture to break data dependencies and dynamically balance workloads.
Our key insight builds on the Markov property of GRWs, which asserts that the next step in a random walk depends solely on the current vertex, independent of prior history. \textit{This property allows us to decompose each GRW query into stateless, fine-grained tasks involving only localized memory access and computation. }
These tasks are inherently independent and can be executed without maintaining global walk state or requiring inter-task synchronization, allowing the architecture to issue massive outstanding memory requests and balance workloads at a much finer granularity.
Based on this insight, we propose \mysys{}, a GRW accelerator that achieves perfectly pipelined execution through two co-design innovations:

\noindent\textbf{Out-of-order Query Execution.} Each GRW traversal hop is further decomposed into several independent, stateless tasks, such as memory access and sampling, that can be issued as soon as their required vertex data becomes available. Tasks from different queries interleave freely, allowing memory stalls in one query to be masked by ready tasks from others (e.g., $Q^1_{s1}$, $Q^1_{s2}$, and $Q^3_{s1}$ share Pipeline 1 in~\Cref{fig:limitations}b), allowing for fully overlapping memory latency with computation.

\noindent\textbf{Workload-Adaptive Scheduling.} To eliminate pipeline bubbles in concurrent GRW query execution, \mysys{} continuously monitors pipeline utilization at cycle-level granularity and dynamically redirects ready tasks to idle modules through a feedback-driven scheduling. For example, in~\Cref{fig:limitations}b, when $Q^5_{s1}$ completes in Pipeline 2 and Pipeline 1 becomes idle, the scheduler immediately dispatches $Q^5_{s2}$ to fill the gap. Such adaptive scheduling eliminates pipeline bubbles and sustains peak throughput even under highly imbalanced GRWs.

\section{RidgeWalker Overview}
\label{sec:overview}

\subsection{Architecture Overview}
\Cref{fig:architecture_overview} illustrates the architecture of {RidgeWalker}, a novel GRW accelerator with perfectly pipelined execution. 
RidgeWalker adopts two tightly integrated components:

\noindent\textbf{A) Asynchronous Pipeline (\S\ref{sec:architecture_desgin}):} To support out-of-order GRW query execution, \mysys{} introduces an asynchronous pipeline design within a scalable architecture. Each asynchronous pipeline independently processes the stages of a GRW step using three key modules: \emph{Row Access}, \emph{Sampling}, and \emph{Column Access}.
The \emph{Row Access} and \emph{Column Access} modules are assigned with independent HBM channels, which avoids inter-channel arbitration and contention, ensuring that random accesses issued by one stage do not interfere with those from another.

Within each pipeline, \mysys{} integrates an asynchronous memory-access engine (see~\Cref{sub:asynchronous_access}) that provides up to 128 outstanding, non-blocking requests. The engine is fully pipelined with an initiation interval of one cycle, enabling it to saturate the outstanding-request capacity of the HBM controller and align memory throughput with the pipeline’s processing rate.
The number of HBM channels assigned to each access engine is selected to match the random-access bandwidth demand of the asynchronous pipeline. For example, a single HBM2 channel can sustain roughly 284 million 64-bit random transactions per second.  A memory-access engine, operating at approximately 300 MHz, is therefore well matched to the capacity of one HBM channel and is able to fully saturate its available random-access bandwidth.

\noindent\textbf{B) Zero-Bubble Query Scheduler (\S\ref{sub:adaptive_workload_balancing}):}
\mysys{} instantiates multiple asynchronous pipelines to fully utilize the available random-access memory bandwidth when executing parallel GRW queries.  To support this parallelism at the data level, the CSR graph is randomly partitioned and distributed across all HBM channels.
Unlike BFS or PageRank, where traversal is tightly tied to partition boundaries and often causes severe load imbalance, GRWs spread their accesses much more evenly. Prior analyses on GRW mixing and meeting times~\cite{kanade2023coalescence,oliveira2019random} show that any imbalance lasts only for a short window of steps, after which access probabilities across partitions become nearly uniform (typically in a few tens of steps for million-scale graphs).

To handle these short-lived fluctuations, \mysys{} allows each walker to be flexibly reassigned at every hop. Multiple parallel pipelines operate asynchronously, and the \emph{Task Router}, implemented with a butterfly interconnect, directs each task to the correct memory channel based on the vertex it needs to access. The zero-bubble scheduler continuously monitors pipeline status and immediately fills open slots with ready tasks, ensuring that temporary bursts do not reduce throughput. Together, the flexible rescheduling and high-throughput routing allow \mysys{} to sustain line-rate execution even under fully random GRW access patterns.

\subsection{Query Execution Flow}
\Cref{fig:architecture_overview}a illustrates how a GRW query executes a single step in \mysys{}, with the corresponding graph memory layout shown in~\Cref{fig:architecture_overview}b.
To match GRW access patterns, the CSR graph is mapped and distributed across HBM channels: the row pointer array is partitioned and assigned to the Row Access channels, while the neighbor lists are shuffled across the Column Access channels in a round-robin manner to reduce the potential of bank conflicts.
Each row pointer entry encodes both the target channel ID and the starting address of the corresponding neighbor list, facilitating direct access.

For a query starting from vertex $V_2$, the \emph{Query Loader} (\circled{1}) fetches it from host memory; the \emph{Scheduler} (\circled{2}) assigns it to Pipeline~2; the \emph{Row Access} module (\circled{3}) retrieves $V_2$'s row pointer. If the neighbor list resides in a different channel, the \emph{Task Router} (\circled{4}) redirects the task accordingly. The \emph{Sampling} module (\circled{5}) selects a neighbor based on a configurable distribution, and the \emph{Column Access} module (\circled{6}) fetches the neighbor and determines the termination of the query.
Each module communicates via shallow FIFOs within the AXI-Stream protocol, enabling backpressure-based flow control. The final list of traversed vertices is collected using the query index upon query termination. This process operates in a streaming-window manner. When the accumulated path length reaches the predefined write granularity, the corresponding paths are sequentially written back to global or host memory that can be used for downstream applications without interrupting GRW execution.

\section{Out-of-order Concurrent Query Execution}
\label{sec:architecture_desgin}

This section describes how the proposed asynchronous pipeline enables out-of-order query execution and resolves data dependencies, and how the asynchronous memory access engine amortizes memory latency to maximize bandwidth utilization under GRW’s highly random access patterns.

\begin{figure}[t!]
  \centering
  \includegraphics[width=\linewidth]{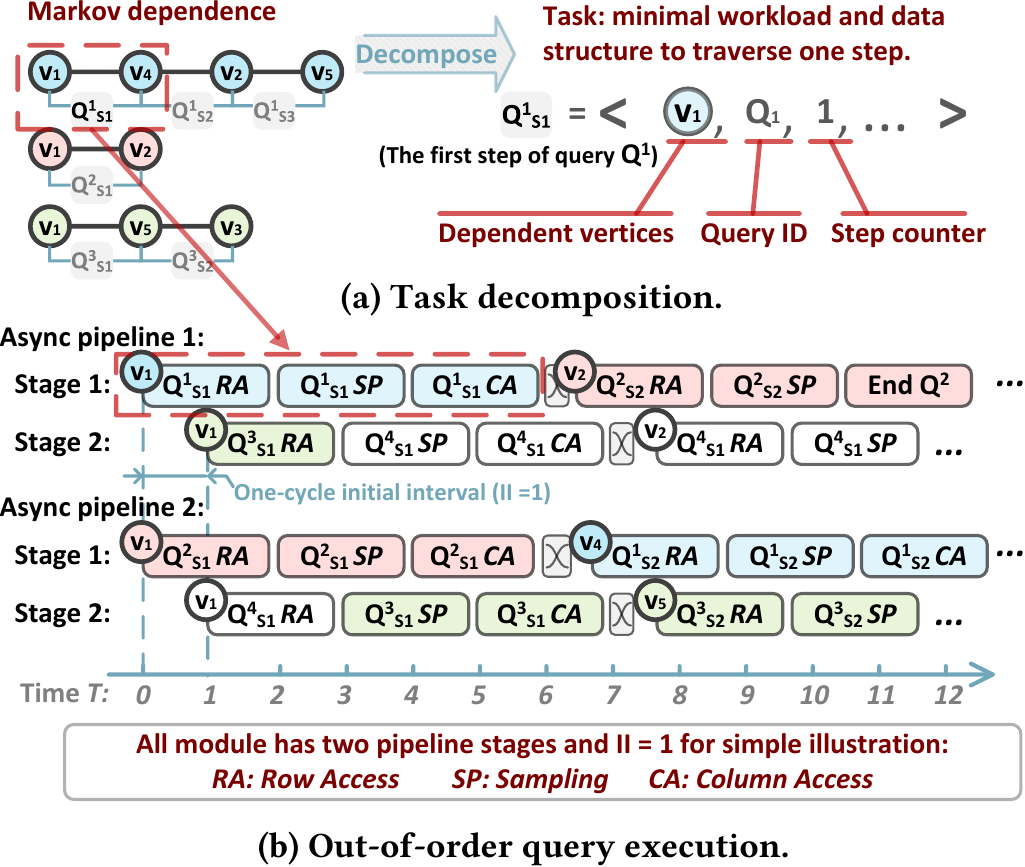}
     \caption{
   Markov-based task decomposition and illustration of out-of-order query execution across two pipelines.}
     \label{fig:decomp}
\end{figure}

\subsection{Decomposing Queries to Stateless Tasks}
\label{sub:asynchronous_grw}
Unlike conventional dataflow architectures~\cite{gao2023fastrw,moreira2020neuronflow,10.1145/3588944}, where execution is strictly governed by data availability, and each module maintains local state to enforce sequential control, \mysys{} eliminates this constraint through Markov-based task decomposition, followed by stateless, out-of-order pipeline execution. Each decomposed task flows independently through the pipeline without requiring additional control logic or global synchronization. Tasks are tagged with a unique query index for result tracking, allowing the system to correctly associate sampled vertices with their corresponding queries. Each module executes based solely on the availability of input tasks and is fully decoupled from the state of other modules. As a result, task execution and rescheduling are non-blocking and interference-free, enabling fully out-of-order query processing.

\Cref{fig:decomp}a illustrates how \mysys{} decomposes a GRW query into minimal, independent tasks. Each task is represented as a tuple \(Q_{s x}^{y} = \langle v_{\text{last}}, \text{ID}_{y}, x, \dots \rangle\), where \(v_{\text{last}}\) denotes the most recently visited vertex (or two vertices for higher-order walks like Node2Vec), \(\text{ID}_{y}\) uniquely identifies the query, and \(x\) tracks the current hop count.
This tuple representation fits within a single pipeline word. Each module consumes the word in one cycle, performs its computation, updates the word, and forwards it to the next module. 
All modules are designed to process one task per cycle, sustaining fully pipelined throughput.

\Cref{fig:decomp}b demonstrates out-of-order execution of decomposed steps across multiple pipelines. Each pipeline comprises a Row-Access (RA), Sampling (SP), and Column-Access (CA) module, and each module is simplified with two pipeline stages and an initiation interval of one cycle for illustration. Benefiting from task independence, they can be dynamically routed across pipelines based on runtime availability, enabling flexible and efficient utilization of all processing resources. For instance, at \(T=6\), Pipeline 1 has already completed \(Q_{s1}^{1}\), and its successor step \(Q_{s2}^{1}\) is immediately dispatched to Pipeline 2, eliminating inter-pipeline blocking. Similarly, the first hop of \(Q^{3}\) begins in Pipeline 1, but subsequent steps are redirected to Pipeline 2 to access graph data mapped to its associated memory. By fully decoupling pipeline modules and removing centralized control via an out-of-order execution model, \mysys{} seamlessly overlaps module-level access with computation.

\begin{figure}[t!]
  \centering
  \includegraphics[width=\linewidth]{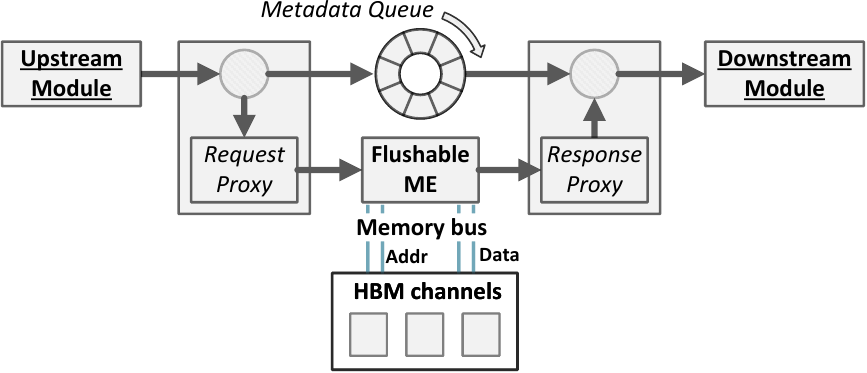}
     \caption{Architecture of asynchronous memory access engine.}
     \label{fig:dme}
\end{figure}

\subsection{Asynchronous Memory Access for Concurrent Execution}
\label{sub:asynchronous_access}

The Asynchronous Access Engine is at the core microarchitectural component underpinning both the \emph{Row Access} and \emph{Column Access} modules. \Cref{fig:dme} presents the architecture. An incoming task first enters the \emph{Request Proxy}, which extracts the target address and accompanying metadata (e.g., vertex index and query ID). The vertex index is translated into a physical address and forwarded to the \emph{Memory Engine}, which issues DRAM accesses at the memory's minimum granularity. Metadata bypasses the data path and is enqueued separately to be reunited with the returned data.

Conventional hardware pipelines stall when input data is unavailable, causing serialization and poor latency hiding. To avoid this, the memory engine operates independently of input readiness signals, allowing access requests to proceed and flush without delay. Each memory request is issued via the AXI protocol and tagged with a unique transaction ID. The associated metadata are enqueued into a BRAM-based \emph{Metadata Queue}, sized to cover the round-trip latency (up to 512 entries, sufficient for 100 cycles at 320~MHz).
Since the AXI protocol ensures in-order responses per transaction ID, our memory engine includes an on-chip buffer supporting up to 64 transaction IDs to reconstruct out-of-order returns. The \emph{Response Proxy} then reassembles the data and metadata into a complete task for downstream modules. This design eliminates pipeline stalls, fully overlaps memory latency, and supports high-throughput, fine-grained random accesses critical for GRWs.

\subsection{Dynamic Query Reassignment Over Parallel Pipelines}
\label{subsec:query_reassign}

\rvh{reassignment}{

To effectively scale GRW execution across a large number of parallel pipelines and memory channels, RidgeWalker must support dynamic per-hop reassignment of queries. The reassignment flexibility in RidgeWalker architecture  arises from two orthogonal dimensions.

First, our Markov-based, per-hop task decomposition ensures that each GRW hop is represented as an independent, stateless task. Because no historical state or dependency chain is carried across hops, every decomposed task can be executed by any pipeline without violating correctness. This provides the foundation for fully flexible, per-hop redistribution.
Second, the hardware overhead of dynamic task routing on FPGA fabrics is minimal. Each decomposed task is compact, no larger than 512 bits, and can be transferred in a single cycle through an AXI-Stream interface. These interfaces require only lightweight resources; even the handshaking FIFO used for flow control can be implemented with a single CLB (32 entries), which already suffices to sustain pipelining among modules. This compact and self-contained task format makes it practical to route and reassign work across many parallel pipelines using only lightweight on-chip interconnects and simple scheduling logic.

Given this flexibility, the key design goal is to maintain balanced execution by considering two factors: (1) each task must be routed to the correct memory channel that stores the adjacency list of its current vertex, and (2) no pipeline should become idle, thereby maximizing throughput.  To achieve this, \mysys{} integrates two architectural mechanisms that leverage task independence. The \emph{Task Router}, implemented as a butterfly interconnect, performs data-aware routing by directing each task to the memory channel holding the required adjacency list. In parallel, the \emph{Zero-Bubble Scheduler} continuously tracks pipeline availability and immediately issues ready tasks into newly freed pipeline slots. Because tasks contain no mutable state, they can be reassigned every cycle without rollback, coordination, or synchronization. As a result, our architecture seamlessly absorbs short-lived workload variations and prevents the formation of bubbles.

Through this combination of stateless task decomposition, data-aware routing, and runtime task dispatch, \mysys{} sustains high utilization across all pipelines and memory channels, while providing the flexibility required for scalable, perfectly pipelined GRW execution.
}

\section{Zero-Bubble Scheduling for Parallel GRW} %
\label{sub:adaptive_workload_balancing}

\Cref{fig:adaptive} illustrates the architecture of the \emph{Zero-Bubble Scheduler}, which dynamically dispatches GRW tasks to pipelines based on real-time execution status. Our design is grounded in a queuing-theoretic model that guarantees full utilization under asynchronous execution, ensuring that tasks are executed without introducing pipeline bubbles.

\subsection{Problem Formulation}
\begin{figure}[t]
  \centering
  \includegraphics[width=1\linewidth]{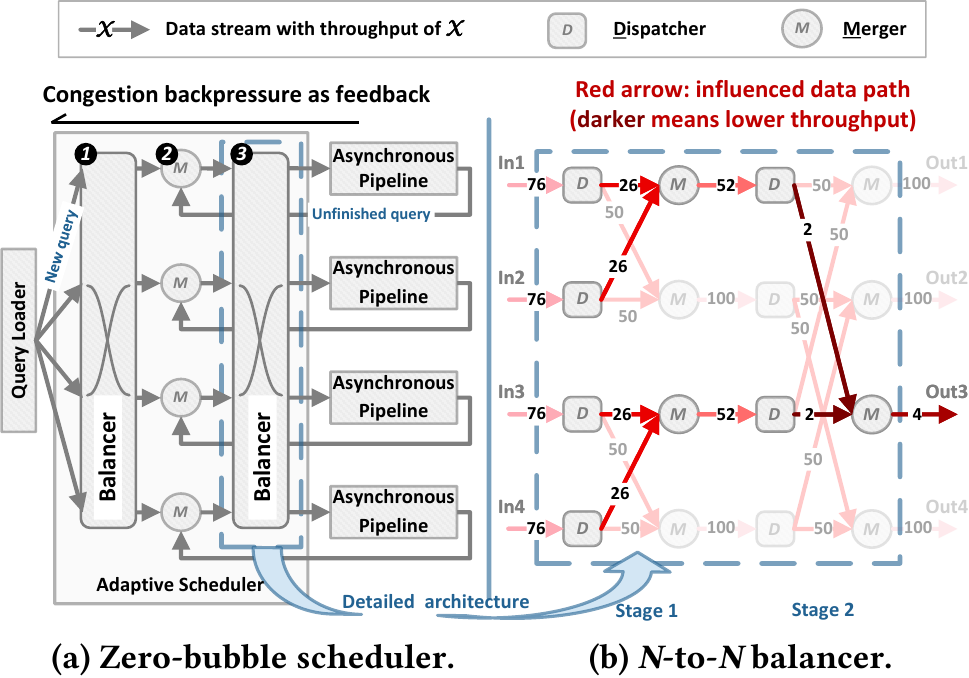}
     \caption{Zero-bubble query scheduler architecture and example of balancing workloads over $\bm{N = 4}$ pipelines.}
     \label{fig:adaptive}
\end{figure}

\noindent\textbf{Hardware View:}
As shown in \Cref{fig:adaptive}, GRW execution is organized as $N$ \emph{asynchronous} pipelines connected through stream FIFOs and an $N$-to-$N$ balancer. Each GRW query is decomposed into a sequence of stateless tasks, but the \emph{per-query execution time is data dependent}: the number of inner-loop iterations (neighbor traversals, rejection retries, early termination, etc.) is unknown a priori. Consequently, different pipelines complete queries at different times, and back-pressure propagates through FIFOs and the balancer. The central challenge is therefore a \emph{hardware scheduling} problem: in every cycle, the scheduler must decide which pipelines to feed using only real-time availability signals (empty/full), such that: i) no pipeline is left idle when work exists (\emph{zero bubbles}), and ii) no subset of pipelines is systematically favored under congestion (\emph{fairness}).

\noindent\textbf{Queuing Model:}
We model the scheduler as a bulk-service queue, specifically an $M/M/1[N]$ system~\cite{dshalalow2023anthology}, to reflect the hardware constraint that the scheduler can issue to multiple pipelines concurrently.
Without loss of generality, we assume that the incoming tasks arrive according to a Poisson process with rate $\lambda$, capturing the stochastic injection of GRW tasks from upstream modules. Task service times are modeled as exponential with rate $\mu$, reflecting the randomized, graph-structure-dependent completion time of GRW work units. The single ``server'' corresponds to the scheduler/balancer logic, which can dispatch up to $N$ tasks in one decision epoch (i.e., one cycle), matching the $N$ parallel pipelines; thus, the maximum batch size in our queuing model is $N$.

\noindent\textbf{Back-pressure and Observation Delay:}
In hardware, the scheduler does not observe instantaneous pipeline idleness; it observes the system through FIFO occupancy and back-pressure signals that are delayed by the interconnect and pipeline depth. We capture this effect with a $C$-cycle delayed observation: at cycle $t$, scheduling decisions are made based on availability signals reflecting the workload state at $t-C$. This delayed feedback, together with stochastic service times, is what creates bubbles in naive static schedules. Our goal is to design a hardware-realizable policy that remains stable (i.e., does not build unbounded queues) for high load and, whenever tasks are present, sustains full utilization of the $N$ pipelines despite the $C$-cycle delay.

\subsection{Buffering Requirement for Zero-Bubble Scheduling}
Given the hardware execution model in \Cref{fig:adaptive}, the scheduler observes pipeline availability through FIFO back-pressure with a finite feedback delay (up to $C$ cycles). Under this delayed observation, insufficient buffering can transiently starve ready pipelines even when work exists upstream, creating bubbles. We therefore apply the queuing-theoretic result of Lu \textit{et al.}~\cite{1203053} (formalized in \Cref{thm:buffer}), which characterizes the minimum queue depth required between a dispatcher (server) and downstream service nodes to maintain a steady state under delayed feedback. In particular, provisioning a queue of depth at least $N + N\mu C$ ensures that the scheduler can continuously supply tasks despite the $C$-cycle delay and stochastic service-time variation. As a result, whenever the system is backlogged, all $N$ pipelines remain fully utilized, achieving zero-bubble scheduling and sustaining peak throughput.

\begin{theorem}[Minimum Buffer Queue Depth]
\label{thm:buffer}
Consider a system with \( N \) independent servers processing tasks, and each server is capable of processing up to \( \mu \) tasks per cycle. Tasks are scheduled from a queue by a scheduler that receives feedback about server availability with a maximum delay of \( C_\text{max} \) time. To ensure that all servers remain fully utilized without idleness, the minimum queue depth \( D \) must satisfy:
\begin{equation}
    D = N + O(\mu C_\text{max}  N).
\end{equation}
\end{theorem}

\subsection{From Queuing Guarantees to Hardware Scheduler Architecture}
\Cref{thm:buffer} provides the buffering condition under which bubbles caused by delayed feedback can be eliminated in principle. We next translate this condition into a hardware-realizable scheduler: a fully pipelined, branch-free dispatch fabric that uses FIFO full/empty signals to route tasks and leverages the derived queue depth to absorb workload imbalance across the $N$ asynchronous pipelines.

The Zero-Bubble Scheduler consists of three core functional modules: \circled1, \circled2, and \circled3 as illustrated in~\Cref{fig:adaptive}a. Together, these modules form a pipelined scheduling datapath designed to achieve high throughput and bubble-free execution. Module~\circled1 serves as the initial task balancer, adaptively distributing newly injected queries from the query loader to available scheduling paths based on pipeline availability. It ensures that incoming tasks are scheduled without stalls introduced by bulk query loading. Module~\circled2 acts as the task merger, combining newly scheduled queries from module~\circled1 with unfinished queries returned from downstream pipelines. This module prioritizes in-flight unfinished queries and enables their immediate redirection into the scheduling pipeline without incurring additional waiting delay. Finally, module~\circled3 functions as the backpressure-aware dispatcher, routing ready-to-run tasks to the appropriate processing pipelines based on real-time availability signals. This multistage coordination enables the scheduler to dynamically balance load and fill execution slots as they are released, thereby eliminating bubbles and sustaining maximum pipeline utilization.

\begin{algorithm}[t]
\SetInd{0em}{1em}
\KwData{\enspace \, $\textit{in}$ : input task stream.
}
\KwResult{
 $\textit{out\_1}: \text{the first output task stream;}$\\
 \enspace \enspace \enspace \enspace \enspace \enspace \enspace \, $\textit{out\_2}: \text{the second output task stream.}$
}
$\textit{last\_selection} = 0$\;

\Do{True}{
    \If{ $\textit{task}=\textit{\textbf{non\_blocking\_read}}()$}
    {
\(
\begin{aligned}
\textit{scode} = \textbf{build\_scode}( &\textit{last\_selection},\\
    &\textit{out\_1}.\textit{\textbf{is\_full}}(), \\
    &\textit{out\_2}.\textit{\textbf{is\_full}}());
\end{aligned}
\)\\
    \Switch{scode}{
            \Case{0b001}{
                \tcp{Both have space; pick not-last-served to alternate (select out\_1).}
            }
            \Case{0b111}{
                \tcp{Both full; block on not-last-served (out\_1) to guarantee fairness.}
            }
            \Case{0b101, 0b100}
            {
                \tcp{Only one channel can accept (out\_2 full); route to out\_1 to avoid stalling.}
                \textit{out\_1}.\textit{\textbf{blocking\_write}}(task)\;
                \textit{last\_selection} = 0\;
                \textbf{break};
            }
            \Other{

                \textit{out\_2}.\textit{\textbf{blocking\_write}}(task)\;
                \textit{last\_selection} = 1\;
                \textbf{break};
                }
    }
    }
}
\caption{Balanced task dispatch.}\label{alg:dispatcher}
\end{algorithm}

\subsubsection{$N$-to-$N$ Task Balancer Design}
\label{subsub:balancer}
Conventional methods for scheduling $N$ tasks across $N$ processors, including the Completely Fair Scheduler~\cite{sun2023neardelayoptimalschedulingbatch,dellacroce2018longestprocessingtimerule,yu2022orlojpredictablyservingunpredictable}, require atomic scheduling to assign all tasks in $O(N \log N)$ time complexity.
The scheduler has to select and commit a global scheduling decision by polling the state of all processors and assigning tasks in a centralized, indivisible operation. Each scheduling action involves $O(\log N)$ complexity per task and introduces strong synchronization dependencies among processors.

In contrast, our task balancer decomposes the scheduling process into independent pairwise comparisons between tasks using lightweight \emph{Dispatcher} and \emph{Merger} modules, each operating in $O(1)$ time. These modules are further scaled with a multistage topology based on a butterfly network, as illustrated in~\Cref{fig:adaptive}b. In each stage, the result of a two-task comparison is propagated forward and participates in the next stage's arbitration to resolve the imbalanced distribution.
Every module is fully pipelined, allowing line-rate scheduling that continuously adapts to the runtime workload distribution.

\begin{algorithm}[t]
\SetInd{0em}{1em}
\KwData{\enspace \, $\textit{in\_1}$ : the first input task stream;\\
\enspace \enspace \enspace \enspace \enspace \enspace \enspace \, $\textit{in\_2}$ : the second input task stream.
}
\KwResult{
 $\textit{out}: \text{output stream.}$
}
$\textit{last\_selection} = 0$\;

\Do{True}{
\(
\begin{aligned}
\textit{scode} = \textbf{build\_scode}( &\textit{last\_selection},\\
    &\textit{in\_1}.\textit{\textbf{is\_empty}}(), \\
    &\textit{in\_2}.\textit{\textbf{is\_empty}}());
\end{aligned}
\)\\
    \Switch{scode}{
            \Case{0b111, 0b110}
            {
               \tcp{Both inputs empty.}
                \textbf{break};

            }

            \Case{0b101, 0b100}{
                \tcp{Only one input has valid data; forward it directly to the output (select in\_1).}
            }

            \Case{0b001}
            {
                \tcp{Both inputs are valid; alternate based on not-last-served to avoid starvation under congestion (select in\_1).}
                \textit{task} = \textit{in\_1}.\textit{\textbf{non\_blocking\_read}}()\;
                \textit{out}.\textit{\textbf{blocking\_write}}(\textit{task})\;
                \textit{last\_selection} = 0\;
                 \textbf{break};
            }
            \Other{
                \textit{task} = \textit{in\_2}.\textit{\textbf{non\_blocking\_read}}()\;
                \textit{out}.\textit{\textbf{blocking\_write}}(\textit{task})\;
                \textit{last\_selection} = 1\;
                 \textbf{break};
            }
    }
}
\caption{Balanced task merge.}\label{alg:merger}
\end{algorithm}

\Cref{fig:adaptive}b exemplifies how the task balancer smooths out downstream load imbalance. Assume the third output channel is limited to 4~pkt/s while the others sustain 100~pkt/s.
In the second stage, each \emph{Dispatcher} feeding the slow channel receives traffic from one fast and one slow path, averaging $(100 + 4)/2 = 52$~pkt/s.
This imbalance is further averaged in the first stage, equalizing all four inputs at 76~pkt/s.
Thus, the multistage routing network spreads local congestion upstream and keeps earlier stages uniformly loaded even when a single downstream channel is throttled.

\subsubsection{Task Dispatcher}
\Cref{alg:dispatcher} details the \emph{Dispatcher} that routes tasks from a single input stream to two independent output channels while honoring back-pressure and preserving fairness. The Dispatcher maintains a one-bit state, \texttt{last\_selection}, initialized in Line~1, to record which output was served most recently. In each iteration (Line~2), it first attempts a \emph{non-blocking} read from the input (Line~3). If no task is available, the Dispatcher simply skips the iteration without stalling upstream.
Once a task is obtained, the Dispatcher constructs a compact three-bit scheduling code, \textit{scode}, using the \texttt{build\_scode()} function (Line~4) by packing \texttt{last\_selection} with the \texttt{is\_full} status of \textit{out\_1} and \textit{out\_2}. Specifically, \texttt{last\_selection} occupies the least significant bit (LSB), while the full flag of \textit{out\_2} is placed in the most significant bit (MSB).

\noindent\textbf{Guaranteeing Balance under Worst-case Congestion.}
This encoding enables a simple decode via a \texttt{switch} (Lines~5-15) rather than complex runtime control branching. In the case that the last task is sent to \textit{out\_2},
when both outputs are available, the Dispatcher selects the \emph{not-last-served} channel to alternate service and balance load at Line~6.
When both outputs are full, it blocks on the \emph{not-last-served} channel to prevent persistent preemption of one side, guaranteeing fairness under worst-case congestion (Line~7).
When only one output can accept new data, the task is routed directly to the available channel (e.g., Line~8).
After each \texttt{blocking\_write}, the Dispatcher updates \texttt{last\_selection} to reflect the chosen output (Lines13–14), ensuring consistent alternation in subsequent iterations. Overall, the Dispatcher is fully pipelined with a one-cycle initiation interval and a fixed latency of two cycles.

\subsubsection{Task Merger}

\Cref{alg:merger} presents the algorithm implemented in \emph{Merger}, which combines two input task streams into a single output stream while maintaining balanced service under back-pressure. The Merger tracks the most recently selected input using a one-bit state, \texttt{last\_selection} (Line~1). In each iteration (Line~2), it constructs a three-bit scheduling code, \textit{scode} (Line~3), in which \texttt{last\_selection} occupies the LSB, and the empty flags of \textit{in\_1} and \textit{in\_2} are stored in the next two bits, respectively.

\noindent\textbf{Guaranteeing Fairness under Worst-case Congestion.}
The scheduling policy is as follows: If exactly one input contains valid data, it forwards that input directly to the output to maximize throughput (e.g., Line~7 and the default case, regardless of the last selection).
When both inputs are valid, the Merger selects the \emph{not-last-served} stream based on \texttt{last\_selection} to alternate input (Line~8). This prevents the output from being continuously occupied by a single input under congestion, avoids starvation, and bounds the worst-case waiting latency for the other stream.
After each successful forward, \texttt{last\_selection} is updated to reflect the chosen input (Lines~11), ensuring persistent balance in subsequent iterations. Similar to the \emph{Dispatcher}, the Merger is fully pipelined with a one-cycle initiation interval and a fixed latency of two cycles.

\subsection{Achieving Perfect Pipelining}
\label{sub:achieving_perfect_pipelining}

We now analyze the feedback delay to apply~\Cref{thm:buffer} to eliminate any pipeline bubbles via scheduling.
In the butterfly topology, each task traverses $\log N$ \emph{Dispatcher} and $\log N$ \emph{Merger} units on the task balancer. Since each unit is fully pipelined with at most two cycles of latency, the total delay through the task balancer is bounded by $2\log N$ cycles. Adding in the round-trip delay from the scheduler to the selected pipeline and back, the total scheduling latency is at most $4\log N$ cycles.
Given that each pipeline sustains an ideal throughput of one GRW step per cycle (i.e., $\mu = 1$),~\Cref{thm:buffer} establishes that a queue depth of $D = N + 4N\log N$ is the minimum required between the scheduler and pipelines to guarantee zero-bubble execution. This corresponds to a FIFO per pipeline with a depth of $1 + 4\log N$, ensuring that no pipeline is idle due to input starvation.

\noindent\textbf{Throughput and Latency Analysis.} Our scheduler is designed to eliminate runtime throughput overhead.
It employs a fully pipelined scheduling datapath with a one-cycle initiation interval, ensuring that both newly injected and redirected queries are concurrently scheduled among multiple processing pipelines at maximum throughput without stalling.
The fixed latency incurred by redirected queries is small (e.g., eight cycles for 16 pipelines), fully overlapped, and amortized with pipeline execution. From the perspective of downstream pipelines, queries remain continuously available, preventing idle cycles due to scheduler delays.

\section{Adaptation to Different GRWs}
\label{sec:different_grw}

\mysys{} is designed to be modular and extensible, supporting a wide range of GRW algorithms used in graph-learning applications. The sampling module communicates via a standard AXI-Stream interface, enabling flexible integration of custom sampling logic. Each instance is paired with ThundeRiNG~\cite{thundering}, an FPGA-optimized, high-throughput random number generator. The task input format encapsulates full query session context, allowing algorithm-specific sampling behavior. Additionally, the graph representation is template-based to support weighted edges and extended metadata. In particular, the row pointer entry ($\textit{RP}_\textit{entry}$) size is configurable at compile time, allowing customization to accommodate auxiliary structures such as alias tables.

\mysys{} supports all commonly used sampling algorithms for both unweighted and weighted graphs, as summarized in~\Cref{tab:sampling}. Uniform sampling is used by URW~\cite{li2015random} and PPR~\cite{hou2021massively}. DeepWalk~\cite{perozzi2014deepwalk} employs alias sampling~\cite{walker1974new,zhang2023efficient}, which requires each neighbor list to maintain an alias table. To accommodate this, the $\textit{RP}_\textit{entry}$ format is extended to 256 bits, storing the alias table pointer and its size. For Node2Vec~\cite{grover2016node2vec}, \mysys{} supports both rejection sampling and reservoir sampling, which are suitable for unweighted and weighted graphs, respectively.

\begin{table}[t]
\centering
\caption{The supported sampling algorithm and GRWs.}
\label{tab:sampling}
\resizebox{\linewidth}{!}{%
\begin{threeparttable}
\begin{tabularx}{1.0\linewidth}{lcYY}
\toprule
\multicolumn{1}{c}{GRWs} &
\multicolumn{1}{c}{Weighted $\graphG?$} &
\multicolumn{1}{c}{Sampling algorithm}&
\multicolumn{1}{c}{ $|\textit{RP}_\textit{entry}|$ } \\

\midrule

\multicolumn{1}{c}{URW~\cite{li2015random}, PPR~\cite{hou2021massively}} &
No &
\multicolumn{1}{l}{Uniform sampling}&64-bit
\\

\multicolumn{1}{c}{DeepWalk~\cite{perozzi2014deepwalk}} &
Yes &
\multicolumn{1}{l}{Alias sampling}&
256-bit
\\

\multicolumn{1}{c}{Node2Vec~\cite{grover2016node2vec}} &
No &
\multicolumn{1}{l}{Rejection sampling}&
64-bit
\\

\multicolumn{1}{c}{Node2Vec~\cite{10.1145/3588944}} &
Yes &
\multicolumn{1}{l}{Reservoir sampling}&
128-bit
\\

\multicolumn{1}{c}{MetaPath~\cite{dong2017metapath2vec}} &
Yes &
\multicolumn{1}{l}{Reservoir sampling}&
128-bit
\\

\bottomrule

\end{tabularx}
\end{threeparttable}
}
\end{table}

To configure the sampling module, \mysys{} exposes memory-mapped \texttt{AXI4-Lite} control registers over PCIe, allowing the host to program algorithm-specific parameters such as the teleport probability \(\alpha\) in PPR or the bias factors \(p\) and \(q\) in Node2Vec. All configuration fields are accessible via lightweight 32-bit register writes. Additionally, sampling modes (e.g., weighted vs. unweighted) can be selected via a mode bit, enabling rapid switching between GRW variants without requiring full resynthesis.

\begin{figure*}[t]
\begin{minipage}[t]{0.31\textwidth}
  \subcaptionbox{DeepWalk on FastRW. \label{fig:fpga_result_dw}}{%
    \includegraphics[height=43mm]{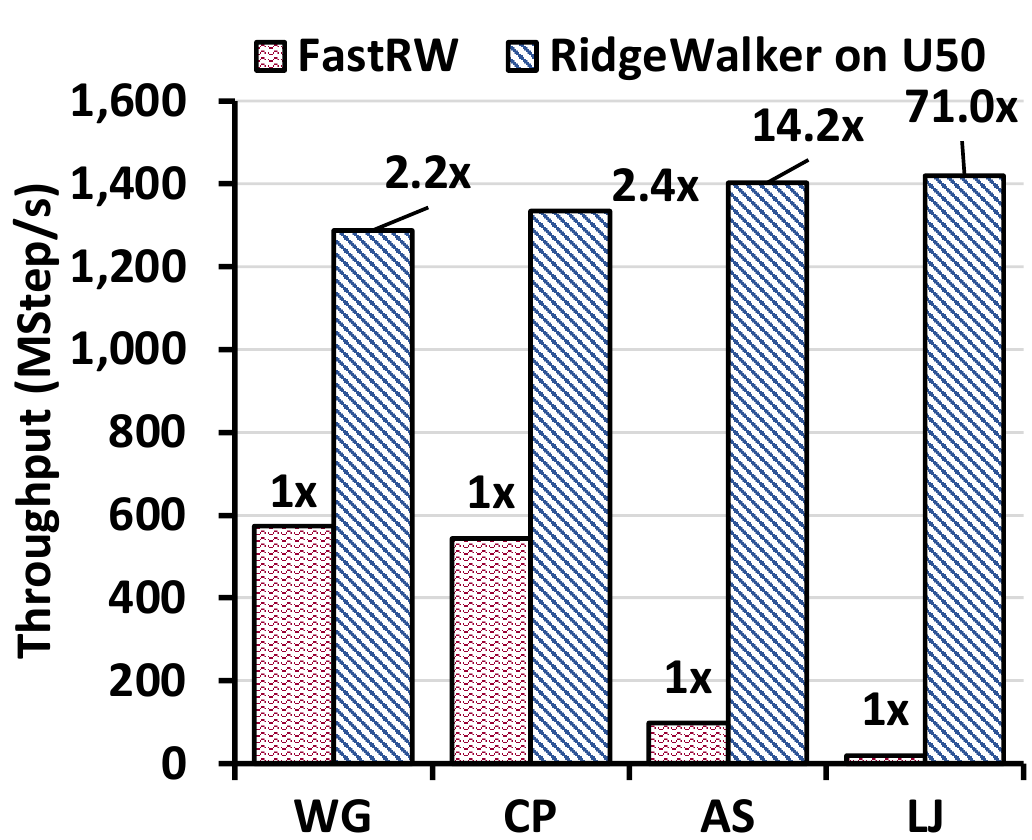}%
  }
\end{minipage}
\hfill
\begin{minipage}[t]{0.25\textwidth}
  \subcaptionbox{PPR and URW on Su et al.~\cite{su2021graph}.\label{fig:fpga_result_ppr}}{%
    \captionsetup{width=\linewidth}
    \includegraphics[height=43mm]{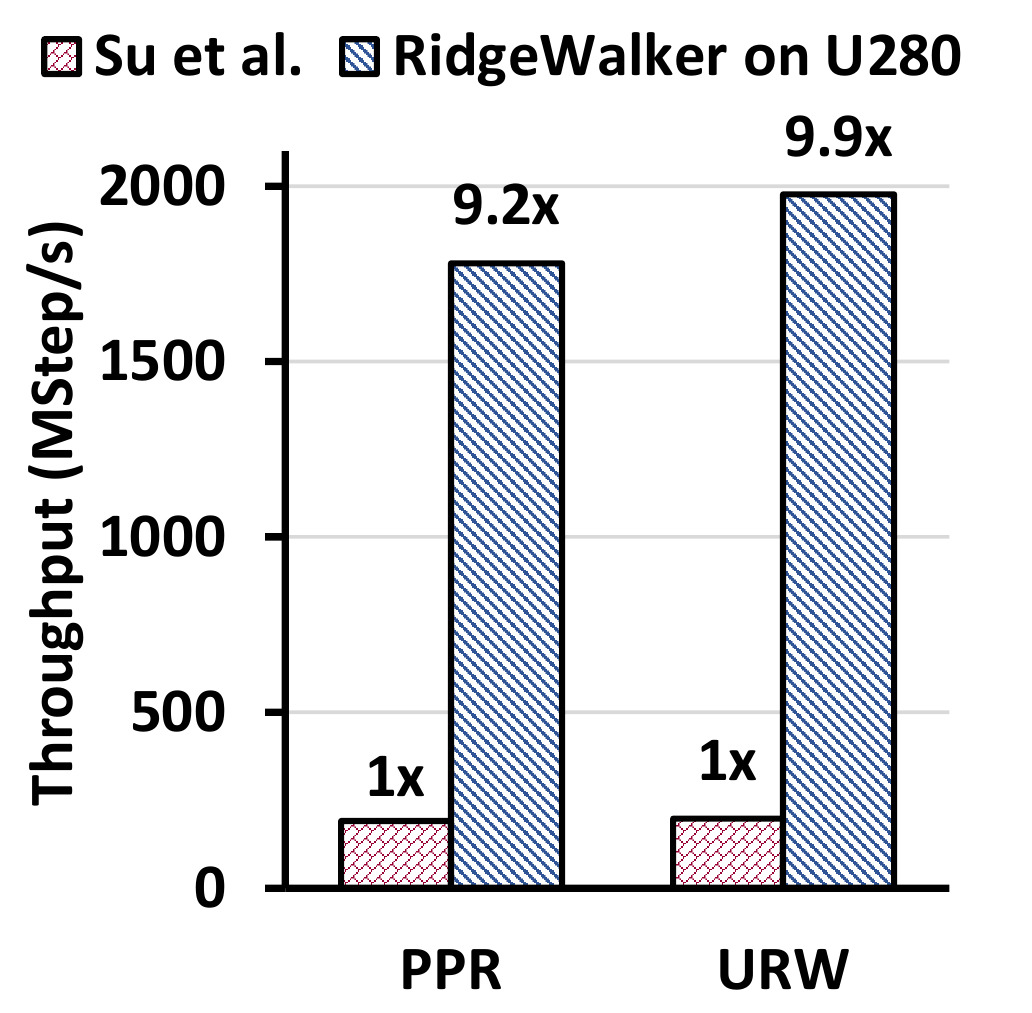}%
  }
\end{minipage}
\hfill
\begin{minipage}[f]{0.40\textwidth}
\vspace{-35mm}
  \subcaptionbox{Node2Vec on LightRW.\label{fig:fpga_result_nv}}{%
   \vspace{-2mm}
    \includegraphics[width=65mm]{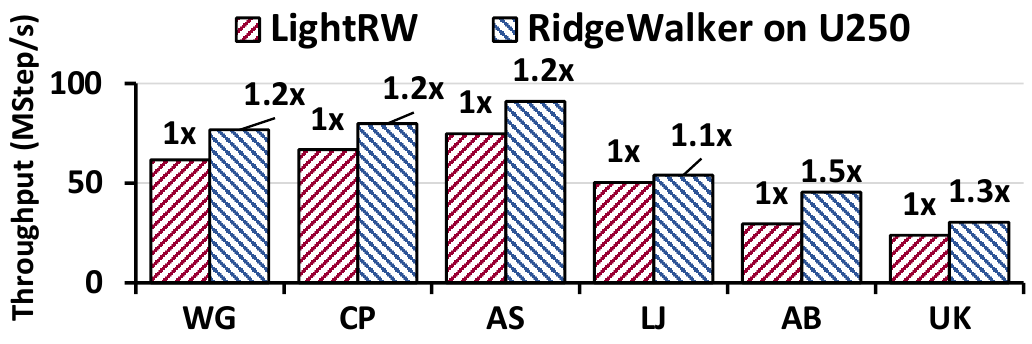}%
  }
  \subcaptionbox{MetaPath on LightRW.\label{fig:fpga_result_mp}}{%
  \vspace{-2mm}
    \includegraphics[width=65mm]{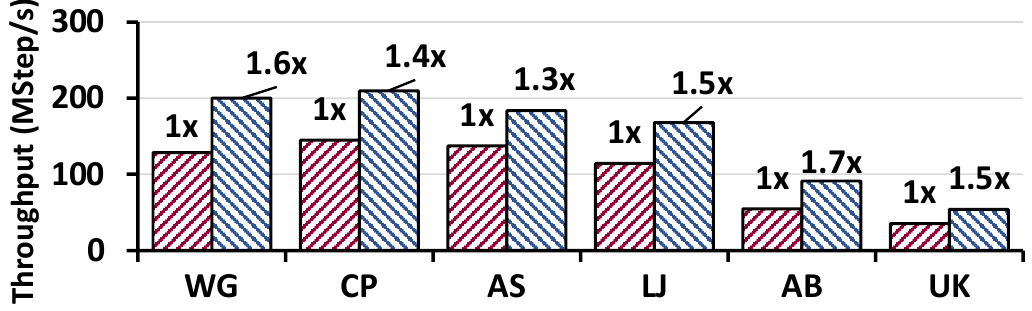}%
  }
\end{minipage}
\caption{Comparison of throughput between \mysys and SOTA FPGA-based GRW accelerators.}
\end{figure*}

\section{Evaluation}
\label{sec:evaluation}
In this section, we present a comprehensive evaluation of \mysys{}, focusing on its asynchronous execution and perfect pipelining. We compare it against state-of-the-art solutions and structure our evaluation around the following objectives:
\begin{enumerate}[leftmargin=*, topsep=2pt,itemsep=-1ex,partopsep=2ex,parsep=1ex]
    \item Evaluate the performance of \mysys{} against state-of-the-art FPGA-based accelerators and highlight the benefits of its perfectly pipelined architectural design~($\mathsection$\ref{sub:comparison_with_state_of_the_art_fpga_imp}).

  \item Compare \mysys{} with high-performance GPU-based solutions, and demonstrate its architectural advantages despite operating under lower available bandwidth~($\mathsection$\ref{sub:comparison_gpu}).

  \item  Quantitatively analyze the effectiveness of the proposed asynchronous access engine and zero-bubble scheduler through microbenchmarking and an ablation study, demonstrating how we address Observation \#1 and \#2 ($\mathsection$\ref{sub:evaluation_on_the_proposed_techniques}).

\end{enumerate}

\subsection{Experimental Setup}
\label{sub:experimental_setup}

\subsubsection{Hardware Setup}
We prototype the \mysys{} design using high-level synthesis (HLS) and mainly evaluate its performance on AMD Alveo U55C FPGA boards. The performance experiments are conducted on a dual-socket AMD EPYC 7V13 128-core CPU server. Since each U55C FPGA supports up to 32 HBM memory channels, and each asynchronous pipeline occupies two HBM channels, we instantiate $32/2 = 16$ asynchronous processing pipelines, following the architecture design in \Cref{sec:overview}. The Zero-Bubble Scheduler is correspondingly configured with 16 outputs connecting to the asynchronous processing pipeline.

\subsubsection{FPGA Baselines}
The design of Su \textit{et~al.}~\cite{su2021graph}, FastRW~\cite{gao2023fastrw} and LightRW~\cite{10.1145/3588944}  are state-of-the-art accelerators for PPR, URW, DeepWalk, Node2Vec and MetaPath, respectively.
To ensure a fair evaluation, \mysys{} is re-synthesised on the same FPGAs used by these baselines as the AMD Alveo U50 (FastRW) and AMD Alveo U250 (LightRW).
Because \mysys{} is architecture-agnostic, porting involved only retiming the memory interface number of memory channels.

\subsubsection{GPU Baseline}
For GPU-based solutions, we benchmark \mysys{} against the state-of-the-art solution, gSampler~\cite{gong2023gsampler}, as gSampler provides the best acceleration performance across all evaluated applications. The experiments are conducted on a server equipped with four NVIDIA H100 PCIe GPUs (H100) features 80 GB of HBM2e memory with a bandwidth of 2093 GB/s running CUDA 12.1.

\begin{table}[t]
\centering
\caption{The evaluated real-world graph datasets.}
\label{tab:graphdataset}
\resizebox{\linewidth}{!}{%
\begin{threeparttable}
\begin{tabularx}{1.07\linewidth}{lXXXcc}
\toprule
\multicolumn{1}{c}{Graphs}&
\multicolumn{1}{c}{$|V|$} &
\multicolumn{1}{c}{$|E|$} &
\multicolumn{1}{c}{Size} &

\multicolumn{1}{c}{Categories} &
\multicolumn{1}{c}{$\delta$}  \\

\midrule

\multicolumn{1}{l}{web-Google (WG)~\cite{snapnets}}&
\multicolumn{1}{c}{$0.9$ M} &
\multicolumn{1}{c}{$5.1$ M} &
\multicolumn{1}{c}{$48$ MB} &
Web &
21
\\

\multicolumn{1}{l}{cit-Patents (CP)~\cite{snapnets}}&
\multicolumn{1}{c}{$3.8$ M} &
\multicolumn{1}{c}{$16.5$ M} &
\multicolumn{1}{c}{$0.2$ GB} &
Citation &
26
\\

\multicolumn{1}{l}{as-Skitter (AS)~\cite{snapnets}}&
\multicolumn{1}{c}{$1.7$ M} &
\multicolumn{1}{c}{$22.2$ M} &
\multicolumn{1}{c}{$0.2$ GB} &
Network &
31
\\

\multicolumn{1}{l}{soc-LiveJournal (LJ)~\cite{snapnets}}&
\multicolumn{1}{c}{$4.9$ M} &
\multicolumn{1}{c}{$69.0$ M} &
\multicolumn{1}{c}{$0.6$ GB} &
Social &
28
\\

\multicolumn{1}{l}{arabic-2005 (AB)~\cite{BoVWFI}}&
\multicolumn{1}{c}{$22.7$ M} &
\multicolumn{1}{c}{$0.6$ B} &
\multicolumn{1}{c}{$5.0$ GB} &
Web &
133
\\

\multicolumn{1}{l}{uk-2005 (UK)~\cite{BoVWFI}}&
\multicolumn{1}{c}{$39.6$ M} &
\multicolumn{1}{c}{$0.8$ B} &
\multicolumn{1}{c}{$6.7$ GB} &
Web &
45
\\

\bottomrule

\end{tabularx}
\end{threeparttable}
}

\end{table}

\subsubsection{GRW Workloads}

\Cref{tab:graphdataset} presents the graph datasets evaluated in our experiments, arranged in ascending order based on the number of edges.
These datasets encompass a variety of real-world graphs, %
including web, network, citation, and social networks.
The fifth column of the table ($\delta$) displays the graph's diameter. That is the longest shortest path between any two vertices.

\noindent\textbf{GRW Algorithms.}
We evaluate four commonly used GRW algorithms from graph database and graph ML application: PPR~\cite{hou2021massively}, URW~\cite{li2015random}, DeepWalk~\cite{perozzi2014deepwalk}, and Node2Vec~\cite{grover2016node2vec}, where DeepWalk and Node2Vec are extensively used in GNN workloads.
To ensure consistency with existing benchmarks~\cite{yang2019knightking,sun2021thunderrw,gong2023gsampler}, we set the query length to 80. For Node2Vec, the parameters are set $p = 2$ and $q = 0.5$, as employed in other state-of-the-art works~\cite{sun2021thunderrw,yang2019knightking,gong2023gsampler}. The edge weights are generated according to the ThunderRW method~\cite{sun2021thunderrw}. For a fair comparison with existing systems, we consistently employ the same state-of-the-art sampling algorithms and configurations, without altering their algorithmic semantics.

\noindent\textbf{Performance Metrics.}
We evaluate GRW execution performance based on effective throughput, which is independent of variations in workload characteristics such as the number of input queries and graph sizes. Throughput is measured in millions of steps per second (MStep/s) and is calculated by dividing the \emph{total count of visited vertices} by the \emph{end-to-end query-processing time}. The processing time is measured as follows: each GRW system is first warmed up and operates as a runtime service with the input graph pre-loaded into main memory. Queries are issued as a continuous stream to emulate real-world application behavior. Each experiment is repeated five times using the same query input, and the median result is reported to reduce the influence of outliers.

\begin{figure*}[t]
\centering
\begin{subfigure}{0.24\linewidth}
\includegraphics[width=\linewidth]{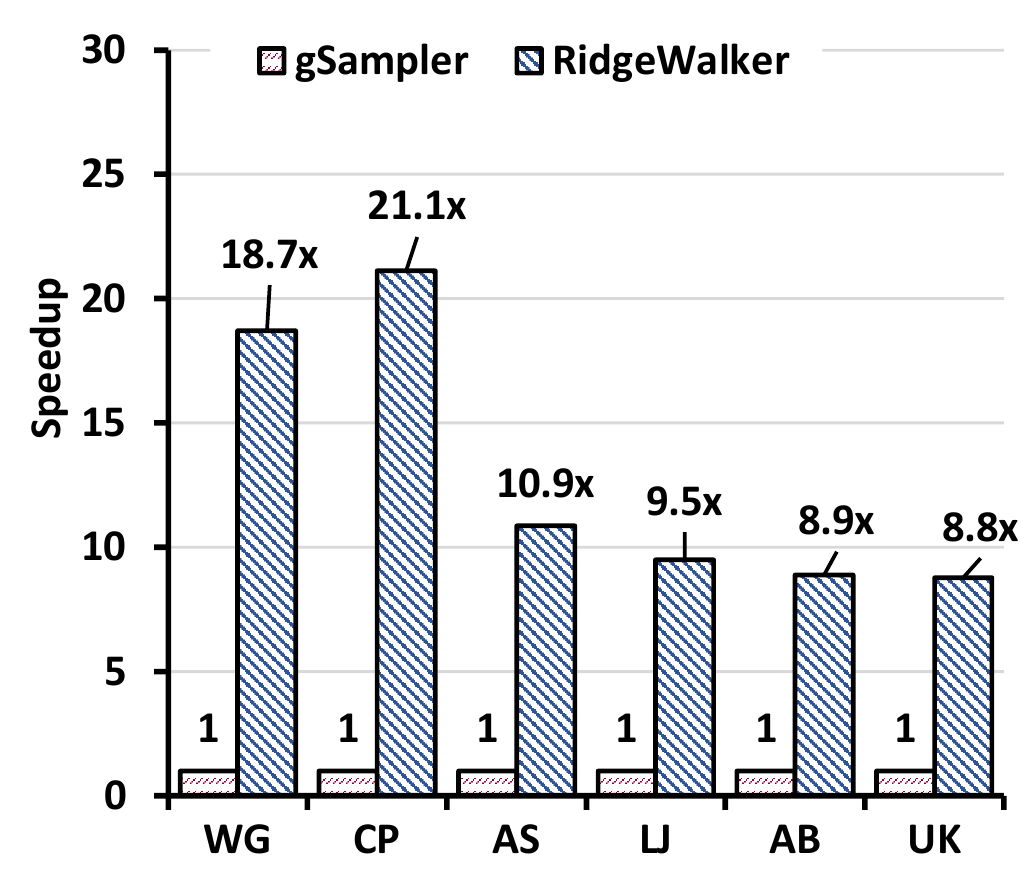}
\vspace{\vrfigsize}
\caption{Personalized PageRank}
\label{fig:gpu_ppr}
\end{subfigure}
\begin{subfigure}{0.24\linewidth}
\includegraphics[width=\linewidth]{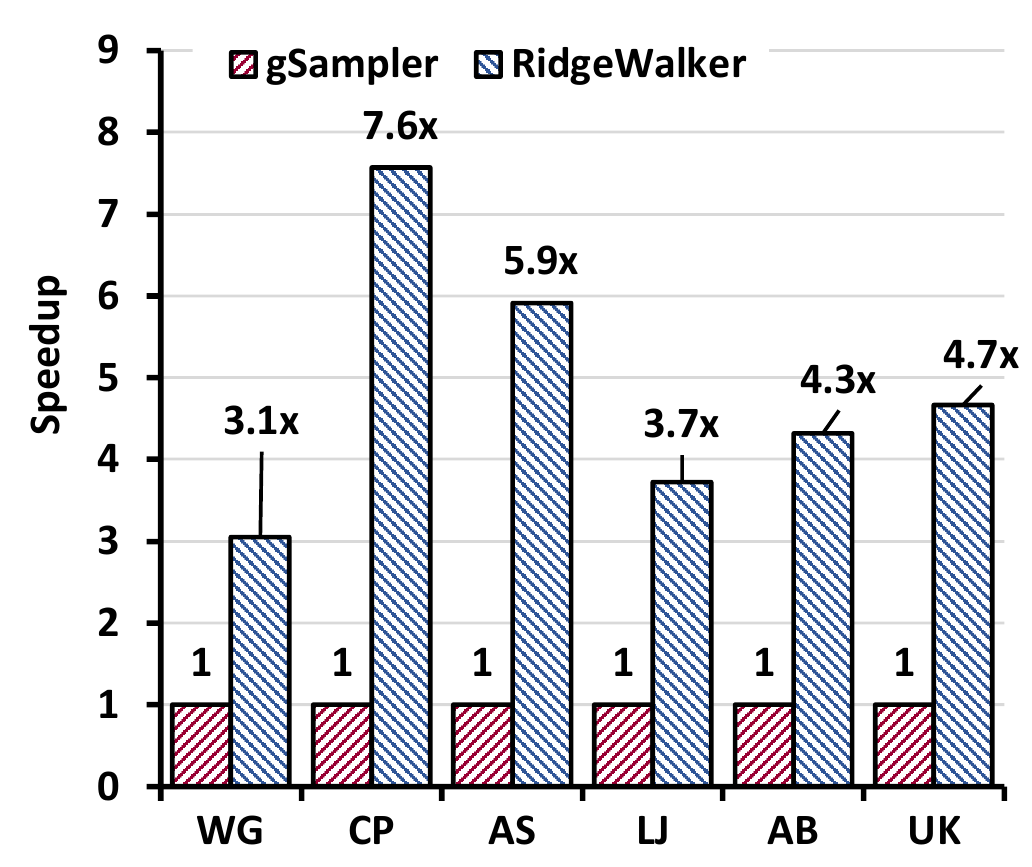}
\vspace{\vrfigsize}
\caption{Uniform random walk}
\label{fig:gpu_urw}
\end{subfigure}
\begin{subfigure}{0.24\linewidth}
\includegraphics[width=\linewidth]{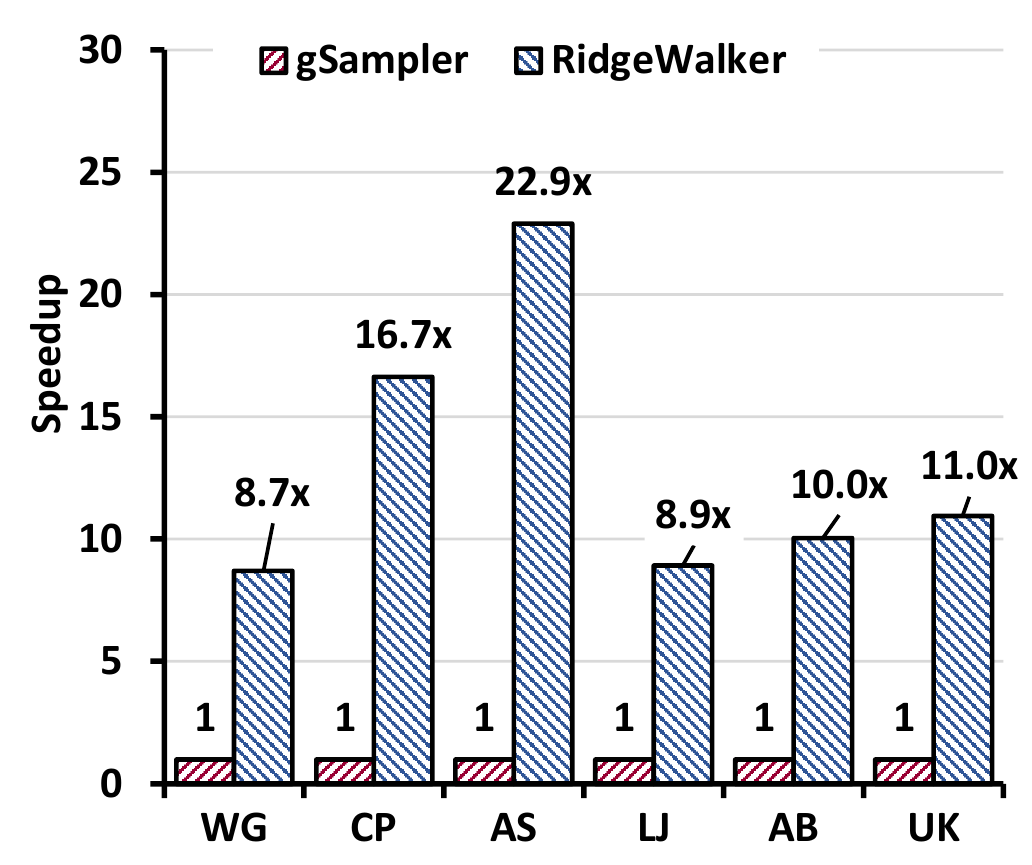}
\vspace{\vrfigsize}
\caption{DeepWalk}
\label{fig:gpu_dp}
\end{subfigure}
\begin{subfigure}{0.24\linewidth}
\includegraphics[width=\linewidth]{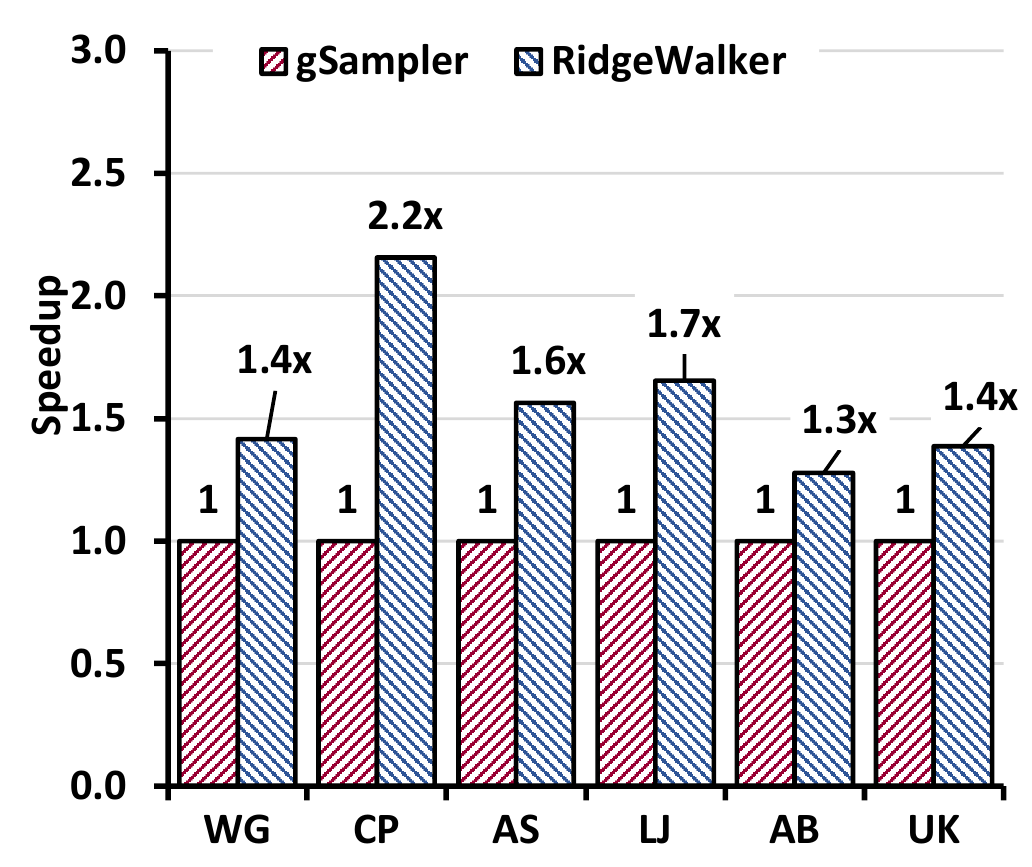}
\vspace{\vrfigsize}
\caption{Node2Vec}
\label{fig:gpu_nv}
\end{subfigure}
\caption{Comparison of normalized throughput to gSampler on four GRW applications.}
\label{fig:comparison_gpu}
\end{figure*}

\subsection{Comparison to SOTA FPGA-based Solutions}
\label{sub:comparison_with_state_of_the_art_fpga_imp}

\Cref{fig:fpga_result_dw} compares the DeepWalk throughput of \mysys{} against FastRW~\cite{gao2023fastrw}, a state-of-the-art GRW accelerator. Since FastRW's code is not publicly available, we implement \mysys{} on the same hardware platform, the Alveo U50 FPGA, and compare its performance using datasets reported in their paper.
\mysys{} consistently outperforms FastRW across all datasets, with speedups increasing with graph size. On the largest dataset (\emph{LJ}), \mysys{} achieves a $70.98\times$ improvement. This is because FastRW relies on caching to store small graphs in limited on-chip memory, which becomes ineffective for large graphs due to GRW’s inherently poor locality. In contrast, \mysys{} is optimized for random access over large graphs residing in HBM.

Even on small graphs like \emph{WG}, which FastRW can cache most of the graph to the on-chip fast memory, \mysys{} also achieves a $2.24\times$ speedup. This is attributed to our out-of-order execution and pipelined pseudo-random number generation using ThundeRiNG~\cite{thundering}, which avoids additional HBM traffic. FastRW, by contrast, pre-generates random numbers on the CPU and has to load them to HBM, consuming the bandwidth that could otherwise be used for graph access.

\Cref{fig:fpga_result_ppr} compares throughput for PPR and URW on the \emph{WG} graph with the accelerator proposed by Su \textit{et al.}~\cite{su2021graph}. Since their code and evaluations are limited to this dataset, we restrict the comparison accordingly. \mysys{} achieves $9.21\times$ and $9.94\times$ higher throughput for PPR and URW, respectively, primarily due to its efficient memory subsystem that fully utilizes HBM bandwidth.

\Cref{fig:fpga_result_nv} compares \mysys{}'s Node2Vec throughput to LightRW~\cite{10.1145/3588944}, using reservoir sampling,  the same sampling method used by LightRW, both implemented on the Alveo U250 FPGA. \mysys{} delivers $1.1\times$–$1.5\times$ higher performance. While both designs are efficient, LightRW uses batched execution, which introduces pipeline stalls. In contrast, \mysys{} adopts fine-grained scheduling, eliminating bubbles and achieving better resource utilization.

A similar trend is observed for MetaPath random walks on weighted graphs. As shown in~\Cref{fig:fpga_result_mp}, \mysys{} achieves a $1.3\times$ to $1.7\times$ speedup over LightRW, due to the higher likelihood of early termination. In MetaPath walks, the next-hop must match a specific type~\cite{sun2021thunderrw}; if no such neighbor exists, the walk ends early. \mysys{} leverages the \emph{Zero-Bubble Scheduler} to maintain high pipeline utilization under this runtime irregularity, consistently outperforming LightRW.

\subsection{Comparison to SOTA GPU-based Solutions} %
\label{sub:comparison_gpu}

\subsubsection{Analysis on Real-world Graphs}\Cref{fig:gpu_ppr} shows that \mysys{} outperforms gSampler on PPR by $8.8\times$–$21.1\times$ across all six graphs, with a peak gain of $21.1\times$ on \texttt{CP}.  The key differentiator is \mysys{}’s {Zero-Bubble Scheduler}. By re-routing ready tasks every cycle, it keeps all pipelines busy even when PPR walks terminate at wildly different lengths, thereby preserving a perfectly filled pipeline and near-peak random access memory bandwidth.
While gSampler relies on \emph{super batching} to reduce kernel‐launch overhead on the GPU, each warp must still wait for the slowest thread. When a random walk ends early, the idle threads induce significant divergence overhead.

\Cref{fig:gpu_urw} reports the speedup of \mysys{} over gSampler on URW. Across all datasets \mysys{} is faster, with gains from $3.1\times$ on \texttt{WG} to $7.6\times$ on \texttt{CP}. The largest improvements on \texttt{CP} and \texttt{AS} arise because our Asynchronous Memory Access Engine saturates random-access bandwidth. The smaller gain on \texttt{WG} reflects its compact size, which fits largely in GPU cache, and the moderate gain on \texttt{LJ} is due to its undirected structure, which reduces workload imbalance. Even so, \mysys{} delivers consistent speedups across graphs.

\Cref{fig:gpu_dp} compares \mysys{} with gSampler on DeepWalk. \mysys{} delivers speedups from $8.7\times$ (\texttt{WG}) to $22.9\times$ (\texttt{AS}); gains exceed $10\times$ on \texttt{CP} and \texttt{AB}. DeepWalk relies on alias sampling, which doubles the number of pseudo-random numbers and increases GPU instruction count, limiting gSampler to just 0.9–2.4 \% of peak bandwidth. \mysys{} fully pipelines sampling and random-number generation, achieving URW-level throughput. These results highlight the importance of pipeline optimization and bandwidth utilization for large-scale graph embedding.

\Cref{fig:gpu_nv} shows the speedup of \mysys{} over gSampler on Node2Vec using rejection sampling~\cite{gong2023gsampler}. Here the speedups are more modest, ranging from $1.28\times$ on \texttt{AB} to $2.16\times$ on \texttt{CP}. This outcome is expected because Node2Vec’s biased walks introduce more structured sequential access on the neighbor list, allowing GPU hardware to capture locality from bulked access.

\begin{figure}[t!]
  \centering
  \includegraphics[width=0.98\linewidth]{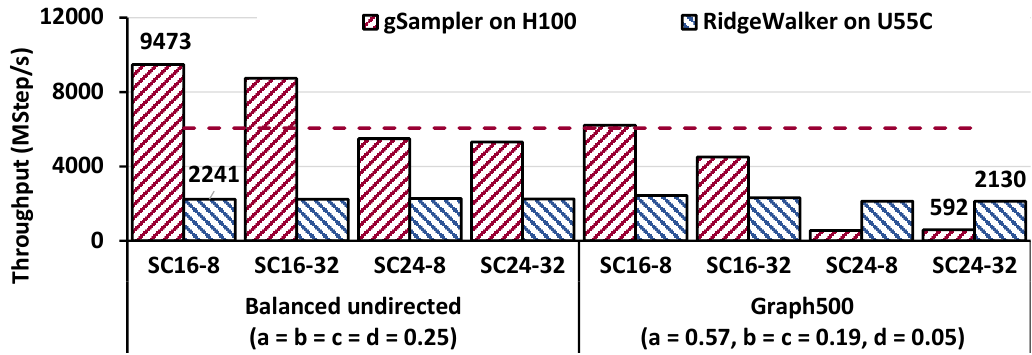}
     \caption{Performance comparison on RMAT graphs under balanced and Graph500 configurations.}
     \label{fig:rmat}
\end{figure}

\subsubsection{Analysis on Synthetic Graphs}
\label{sub:analysis_on_synthetic_graph}
To investigate and further breakdown the performance benefits of \mysys{}, we compare it against DeepWalk in gSampler on H100 GPU using synthetic RMAT~\cite{chakrabarti2004r} graphs with varying size and density. Two initiator configurations are evaluated: the balanced undirected setup (RMAT probability distribution is set as a=b=c=d=0.25) and the Graph500 configuration (a=0.57, b=c=0.19, d=0.05)~\cite{murphy2010introducing}. Each graph is labeled as SC$X$-$Y$, where $X$ represents the scale factor and $Y$ denotes the edge factor (e.g., SC24-32 corresponds to scale 24 and edge factor 32).

The red dashed line in~\Cref{fig:rmat} represents the benchmarked upper bound of GRW throughput on the H100 GPU, derived from its measured random-access memory bandwidth (excluding cache effects). derived from the random-access bandwidth benchmark~\cite{lutz:sigmod:2020}.
On SC24 balanced graphs, gSampler’s throughput closely approaches this line, showing that GPU execution achieves near-peak random-access efficiency when accesses are evenly distributed.

Although the GPU achieves high absolute throughput on balanced RMAT graphs, the picture changes dramatically under the skewed Graph500 configuration. Graph500 introduces a strong structural imbalance, causing traversing lengths to vary significantly across queries. Under the SIMT execution model, GPU warps execute in lockstep, so threads that finish early must stall until the longest walk completes. This leads to heavy divergence and underutilization, reducing throughput by more than an order of magnitude compared with the balanced case. Increasing the number of queries does not help, as millions of concurrent walks are already issued and all SMs are fully saturated.

\mysys{} maintains consistently high throughput across all RMAT configurations. Our stateless task decomposition and zero-bubble scheduling enable fully independent per-hop execution, allowing short and long walks to proceed without blocking one another. As a result, \mysys{} continues to deliver around 2,130 MSteps/s even on the heavily skewed graphs, demonstrating that architectural tolerance to workload imbalance can outweigh raw bandwidth advantages. This highlights a key strength of our design: \mysys{} converts irregular GRW workloads into perfectly pipelined execution, allowing it to fully exploit the hardware potential.

\begin{figure}[t]
\centering
\includegraphics[width=\linewidth]{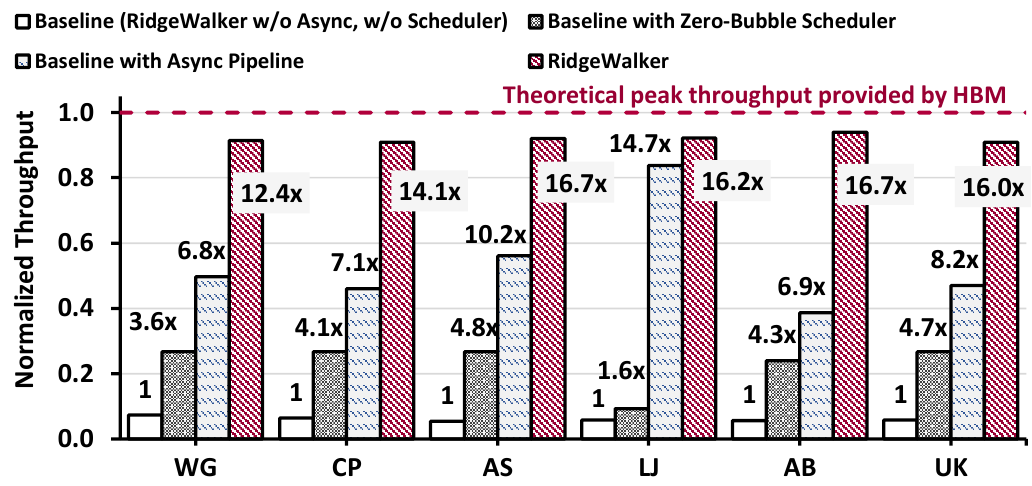}%
\caption{Breakdown of performance gains from the Asynchronous Pipeline and Zero-Bubble Scheduler\label{fig:breakdown_ana}}
\end{figure}

\subsection{Breakdown Evaluation}
\label{sub:evaluation_on_the_proposed_techniques}

\rvh{breakdown}{

\Cref{fig:breakdown_ana} reports the normalized throughput across all real-world graphs, with all values shown relative to the theoretical HBM-supported throughput (indicated by the red dashed line and computed using~\Cref{eq:dram}). Each bar reflects both the achieved performance and its speedup over our breakdown baseline. The baseline retains the overall \mysys{} architecture but disables both asynchronous execution and dynamic scheduling: queries are statically bound to fixed pipelines,
memory requests are issued directly to HBM via a standard AXI interface,
and execution proceeds in bulk-synchronous batches without early-termination handling,
similar to LightRW~\cite{10.1145/3588944} and FastRW~\cite{gao2023fastrw}.
We then progressively enable individual optimizations to isolate their contributions.

{First}, enabling only the zero-bubble scheduler allows early-terminating queries
to be dynamically redistributed across pipelines.
This eliminates pipeline bubbles and improves performance by
$1.6\times$--$4.8\times$.
The improvement is small on \emph{LJ}, whose undirected structure results in
very few early terminations.
Given that nearly $80\%$ of all 418 graphs from KONECT~\cite{kunegis2013konect} are directed,
the zero-bubble scheduler is crucial for sustaining high utilization in practical GRW workloads,
directly addressing Observation~\#2, mitigating the substantial performance loss caused by early termination in GRW queries.

{Second}, enabling the asynchronous pipeline and the asynchronous memory-access
engine, while keeping the scheduler disabled, isolates the effect of latency hiding and
out-of-order task execution.
This design improves baseline performance by $6.8\times$ to $14.7\times$,
demonstrating that decoupled memory access and per-hop task independence are essential for mitigating memory-latency bottlenecks in GRW execution. It addresses Observation~\#1 over prior work by efficiently amortizing pointer-chasing latency across multiple concurrent queries.

{Finally}, enabling both optimizations together yields perfectly pipelined execution
across massive numbers of concurrent queries.
Under this configuration, \mysys{} achieves up to $16.7\times$ speedup over the baseline and
reaches up to $88\%$ of the theoretical HBM random-access peak.

}

\begin{table}[t]
\centering
\caption{Average URW throughput of all graph datasets across FPGA with different memory configurations}
\label{tab:differentfpga}
\begin{threeparttable}
\begin{tabularx}{0.94\linewidth}{lcccc}
\toprule
\multicolumn{1}{c}{\multirow{1}{*}{{\text{}}}}
& \multirow{1}{*}{{U250}}
& \multirow{1}{*}{{VCK5000}}
& \multirow{1}{*}{{U50}}
& \multirow{1}{*}{{U55C}}
\\

\midrule
\multicolumn{1}{c}{Memory type} & DDR4   &  DDR4-NoC  & HBM2 & HBM2 \\
\multicolumn{1}{c}{Bandwidth (GB/s)} & 77  & 102 & 316   & 460  \\
\multicolumn{1}{c}{\# of memory channel} & 4  & 4 & 32   & 32  \\
\multicolumn{1}{c}{Throughput (MStep/s)} & 258  & 202 &  1463 & 2098 \\
\multicolumn{1}{c}{BW utilization} & 81\%  & 87\% & 88\%  & 88\% \\

\bottomrule
\end{tabularx}%
\end{threeparttable}
\end{table}

\subsection{Support on Different FPGAs}
\label{sub:support_on_different_fpgas}

To demonstrate the generalizability of \mysys{}’s architecture, we evaluate its performance on four FPGA platforms: U250, VCK5000, U50, and U55C. These devices vary in available memory bandwidth and the number of independent memory channels, as summarized in the second and third rows of~\Cref{tab:differentfpga}. We measure both the throughput (fourth row) and random-access bandwidth utilization (fifth row) of URW across all graph datasets listed in~\Cref{tab:graphdataset}, and report their average values.
The Versal VCK5000 includes a hardened network-on-chip (NoC) memory subsystem with four DDR4 channels, providing up to 102~GB/s aggregate bandwidth. To support irregular access patterns, we disable default NoC channel interleaving, typically optimized for sequential workloads, and deploy \mysys{} with four processing pipelines. Overall, \mysys{} sustains bandwidth utilization from 81\% to 88\% across all evaluated FPGA platforms. \mysys{} requires only independent memory channels that support the standard AXI protocol, facilitating portability across different FPGAs.

\begin{table}[t]
\centering
\caption{The consumption of hardware resource (percentage) and frequency (MHz) of different GRWs on U55C FPGA.}
\label{tab:resource}
\resizebox{\linewidth}{!}{%
\begin{threeparttable}
\begin{tabularx}{1\linewidth}{lYYYYY}
\toprule
\multicolumn{1}{c|}{\multirow{1}{*}{{\text{App.}}}}
& \multirow{1}{*}{{LUTs}}
& \multirow{1}{*}{{REGs}}
& \multirow{1}{*}{{BRAMs}}
& \multirow{1}{*}{{DSPs}}
& \multirow{1}{*}{{Frequency}}
\\

\midrule

\multicolumn{1}{c|}{PPR} & $61.1\%$  & $29.8\%$ & $19.5\%$ & $2.2\%$ &320MHz \\
\multicolumn{1}{c|}{URW} & $50.1\%$  & $24.0\%$ & $19.5\%$ & $2.2\%$ &320MHz \\
\multicolumn{1}{c|}{DeepWalk} & $67.5\%$  & $32.3\%$ & $39.1\%$ & $4.4\%$ &320MHz \\
\multicolumn{1}{c|}{Node2Vec} & $79.1\%$  & $41.6\%$ & $36.0\%$ & $7.3\%$ &320MHz \\

\bottomrule
\end{tabularx}%
\end{threeparttable}
}
\end{table}

\subsection{Resource Utilization and Frequency Optimization}
\label{sub:resource_utilization_and_frequency_optimization}

\Cref{tab:resource} reports the resource utilization and operating frequency of \mysys{} for four GRW kernels implemented on the U55C FPGA. Because GRWs are fundamentally memory bound, the design focuses on saturating the available random access memory bandwidth and channels rather than on exhausting logic resources, leaving ample capacity for additional, downstream accelerators. Resource usage varies with the sampling method and the size of each row-pointer entry (\textit{RP}\textsubscript{entry}, see~\Cref{sec:different_grw}). Benefiting from the asynchronous execution model, operators and modules are decoupled, simplifying timing closure and supporting a frequency up to 320 MHz.

\mysys{} implements shallow FIFOs with LUTs, streamlining placement and routing for high-frequency. BRAM is used for deeper buffers, as each block can store up to 512 entries. Two components rely on BRAM-based FIFOs: (1) a 128-entry metadata queue in the asynchronous memory-access engine, sized to absorb the HBM round-trip latency, and (2) 65-entry FIFOs between the scheduler and the pipelines, which maintain a steady query processing and prevent stalls.

We further optimize the zero-bubble scheduler to prevent the butterfly interconnect from becoming a critical path.  We inserted registers to break long combinational logic paths, and the entire module is designed as a free-running module without global control signals such as start/stop triggers. This eliminates high-fanout broadcast logic, improving timing for higher frequency. Independent profiling on the U55C FPGA shows that the scheduler operates at up to 450~MHz
while consuming only 1.8\% of available LUTs, demonstrating its potential scalability beyond 32 HBM channels.

\section{Related Work and Discussion}
\label{sec:related_work}

\noindent\textbf{ASIC and In-Memory Designs.}
\label{subsec:asic-rw}
Several works tackle the memory-bound nature of random walks by bringing computation closer to data. FlashWalker~\cite{niu2022flashwalker} embeds walk logic within solid-state drives, leveraging NAND-level parallelism to eliminate host-device transfers. Other efforts explore {\em processing-in-memory} (PIM) using emerging technologies such as ReRAM, enabling graph processing directly within memory arrays to reduce data movement and latency~\cite{choudhury2023accelerating}. However, current PIM designs support only very small graph sizes and remain limited in scalability.
While ASIC and PIM approaches offer performance potential, they face challenges in adaptability to evolving algorithms and long development cycles. FPGA-based solutions are flexible to deploy and adaptive to the fast evolution of GRW algorithms.

\noindent\textbf{Supercomputing Architectures for Graph Processing.}
General-purpose multithreaded architectures such as the Cray XMT (originally Tera MTA)~\cite{bader2006designing,mizell2009early} and Lucata Emu~\cite{hein2020programming,smith2022concurrentgraphquerieslucata} have been explored for irregular graph processing workloads, including BFS and SSSP. These systems tolerate memory latency by maintaining hundreds of hardware thread contexts and issuing instructions from any ready thread at each time.
\mysys{} takes a different approach by adopting a {\em domain-specific architecture} tailored for GRWs. Leveraging the Markov property, it allows query tasks to flow directly through the pipeline as fine-grained stateless units, enabling out-of-order execution without the need to maintain thread contexts or global walk state in memory. Furthermore, \mysys{} employs a scheduler that is formally grounded in queueing theory, sustaining near-optimal throughput even under highly diverse and imbalanced GRW workloads.

\noindent\textbf{Sparse Data Structure Fetchers.}
Widx~\cite{kocberber2013meet} accelerates database hash-index lookups using a stateful design that offloads pointer chasing and key hashing to programmable on-chip units with per-key state in dispatcher tables and local caches.
Fifer~\cite{nguyen2021fifer} employs coarse-grained time-multiplexed reconfiguration to tolerate memory latency and balance irregular workloads.
TMU~\cite{zhou2025tensor} targets data-movement–intensive tensor operators using reconfigurable address abstractions for coarse- and fine-grained data rearrangement.
Terminus~\cite{lee2024terminus} generalizes sparse data structure acceleration with per-operation and per-partition state to support fine-grained updates on hash tables and trees.
Aurochs~\cite{vilim2021aurochs} extends dataflow accelerators to execute irregular structures via stateful fine-grained hardware threading.
These designs are mainly stateful, maintaining per-query or per-task context in registers, caches, or buffers to overlap memory access and computation.
In contrast, \mysys{} adopts a stateless task decomposition model based on the Markov property, enabling flexible scheduling and massive parallelism. The architecture is specialized for GRW workloads, achieving near-optimal random-access throughput and high efficiency for emerging large-scale graph applications.

\noindent\textbf{Discussion.}
\mysys{} employs standard architecture primitives rather than FPGA-specific features, hence it can be easily ported to ASICs and complements existing logic. Furthermore, our perfect pipelining strategy generalizes to other probabilistic workloads, such as Bayesian networks and Monte Carlo Markov Chain applications, where runtime dependencies and random access latency critically affect performance.

\section{Conclusion}
\label{sec:conclusion}

This paper introduces RidgeWalker, a perfect pipelined FPGA-based accelerator
that leverages the Markovian property of GRWs to decompose queries
into stateless tasks. By combining several asynchronous pipelines with the proposed zero-bubble scheduler on FPGAs, RidgeWalker maximizes parallelism and achieves 88\% runtime utilization of the theoretical random-access memory bandwidth. \mysys{} delivers up to $71.0\times$ and $22.9\times$ speedup over state-of-the-art FPGA and GPU solutions, respectively. Moreover, RidgeWalker’s modular design supports diverse GRW algorithms, offering a scalable and efficient architecture for high-performance GRWs.

\section*{Acknowledgements}

This research/project is supported by the Ministry of Education AcRF Tier
2 grant (No. MOE-T2EP20224-0020) and Tier 1 grant (No. T1 251RES2315) in Singapore, the Google South \& Southeast Asia Research Award
2025, and the Natural Sciences and Engineering Research Council of Canada (NSERC) Discovery Grant. We also thank the AMD Heterogeneous Accelerated Compute Clusters (HACC) program~\cite{hacc}
for the generous hardware donation. Yao Chen from the National University of Singapore is the corresponding author.

\bibliographystyle{IEEEtran}
\bibliography{reference}

@inproceedings{lee2024terminus,
  title={Terminus: A Programmable Accelerator for Read and Update Operations on Sparse Data Structures},
  author={Lee, Hyun Ryong and Sanchez, Daniel},
  booktitle={2024 57th IEEE/ACM International Symposium on Microarchitecture (MICRO)},
  pages={1233--1246},
  year={2024},
  organization={IEEE}
}

@inproceedings{vilim2021aurochs,
  title={Aurochs: An architecture for dataflow threads},
  author={Vilim, Matthew and Rucker, Alexander and Olukotun, Kunle},
  booktitle={2021 ACM/IEEE 48th Annual International Symposium on Computer Architecture (ISCA)},
  pages={402--415},
  year={2021},
  organization={IEEE}
}

@inproceedings{nguyen2021fifer,
  title={Fifer: Practical acceleration of irregular applications on reconfigurable architectures},
  author={Nguyen, Quan M and Sanchez, Daniel},
  booktitle={MICRO-54: 54th Annual IEEE/ACM International Symposium on Microarchitecture},
  pages={1064--1077},
  year={2021}
}

@article{zhou2025tensor,
  title={Tensor Manipulation Unit (TMU): Reconfigurable, Near-Memory Tensor Manipulation for High-Throughput AI SoC},
  author={Zhou, Weiyu and Wang, Zheng and Chen, Chao and Li, Yike and Yang, Yongkui and Wu, Zhuoyu and Chattopadhyay, Anupam},
  journal={arXiv preprint arXiv:2506.14364},
  year={2025}
}

@inproceedings{kocberber2013meet,
  title={Meet the walkers: Accelerating index traversals for in-memory databases},
  author={Kocberber, Onur and Grot, Boris and Picorel, Javier and Falsafi, Babak and Lim, Kevin and Ranganathan, Parthasarathy},
  booktitle={Proceedings of the 46th Annual IEEE/ACM International Symposium on Microarchitecture},
  pages={468--479},
  year={2013}
}

@article{kanade2023coalescence,
  title={On coalescence time in graphs: When is coalescing as fast as meeting?},
  author={Kanade, Varun and Mallmann-Trenn, Frederik and Sauerwald, Thomas},
  journal={ACM Transactions on Algorithms},
  volume={19},
  number={2},
  pages={1--46},
  year={2023},
  publisher={ACM New York, NY}
}

@inproceedings{oliveira2019random,
  title={Random walks on graphs: new bounds on hitting, meeting, coalescing and returning},
  author={Oliveira, Roberto I and Peres, Yuval},
  booktitle={2019 Proceedings of the Sixteenth Workshop on Analytic Algorithmics and Combinatorics (ANALCO)},
  pages={119--126},
  year={2019},
  organization={SIAM}
}

@article{murphy2010introducing,
  title={Introducing the graph 500},
  author={Murphy, Richard C and Wheeler, Kyle B and Barrett, Brian W and Ang, James A},
  journal={Cray Users Group (CUG)},
  volume={19},
  number={45-74},
  pages={22},
  year={2010}
}

@article{wang2023optimizing,
  title={Optimizing GPU-based graph sampling and random walk for efficiency and scalability},
  author={Wang, Pengyu and Xu, Cheng and Li, Chao and Wang, Jing and Wang, Taolei and Zhang, Lu and Hou, Xiaofeng and Guo, Minyi},
  journal={IEEE Transactions on Computers},
  volume={72},
  number={9},
  pages={2508--2521},
  year={2023},
  publisher={IEEE}
}

@article{lambiotte2015random,
  title={Random walks, Markov processes and the multiscale modular organization of complex networks},
  author={Lambiotte, Renaud and Delvenne, Jean-Charles and Barahona, Mauricio},
  journal={IEEE Transactions on Network Science and Engineering},
  volume={1},
  number={2},
  pages={76--90},
  year={2015},
  publisher={IEEE}
}

@inproceedings{mizell2009early,
  title={Early experiences with large-scale Cray XMT systems},
  author={Mizell, David and Maschhoff, Kristyn},
  booktitle={2009 IEEE International Symposium on Parallel \& Distributed Processing},
  pages={1--9},
  year={2009},
  organization={IEEE}
}

@misc{smith2022concurrentgraphquerieslucata,
      title={Concurrent Graph Queries on the Lucata Pathfinder},
      author={Emory Smith and Shannon Kuntz and Jason Riedy and Martin Deneroff},
      year={2022},
      eprint={2209.11889},
      archivePrefix={arXiv},
      primaryClass={cs.DC},
      url={https://arxiv.org/abs/2209.11889},
}

@inproceedings{bader2006designing,
  title={Designing multithreaded algorithms for breadth-first search and st-connectivity on the Cray MTA-2},
  author={Bader, David A and Madduri, Kamesh},
  booktitle={2006 International Conference on Parallel Processing (ICPP'06)},
  pages={523--530},
  year={2006},
  organization={IEEE}
}

@article{hein2020programming,
  title={Programming strategies for irregular algorithms on the emu chick},
  author={Hein, Eric R and Eswar, Srinivas and Ya{\c{s}}ar, Abdurrahman and Li, Jiajia and Young, Jeffrey S and Conte, Thomas M and {\c{C}}ataly{\"u}rek, {\"U}mit V and Vuduc, Richard and Riedy, Jason and U{\c{c}}ar, Bora},
  journal={ACM Transactions on Parallel Computing (TOPC)},
  volume={7},
  number={4},
  pages={1--25},
  year={2020},
  publisher={ACM New York, NY, USA}
}

@article{jaiyeobaswift,
  title={Swift: A Multi-FPGA Framework for Scaling Up Accelerated Graph Analytics},
  author={Jaiyeoba, Oluwole and Mughrabi, Abdullah T and Baradaran, Morteza and Gul, Beenish and Skadron, Kevin},
  booktitle={International Conference on Field Programmable Technology},
  year={2024}
}

@inproceedings{jaiyeoba2023acts,
  title={Acts: a near-memory FPGA graph processing framework},
  author={Jaiyeoba, Wole and Elyasi, Nima and Choi, Changho and Skadron, Kevin},
  booktitle={Proceedings of the 2023 ACM/SIGDA International Symposium on Field Programmable Gate Arrays},
  pages={79--89},
  year={2023}
}

@article{jaiyeoba2024dynamic,
  title={Dynamic-ACTS-A Dynamic Graph Analytics Accelerator For HBM-Enabled FPGAs},
  author={Jaiyeoba, Oluwole and Skadron, Kevin},
  journal={ACM Transactions on Reconfigurable Technology and Systems},
  volume={17},
  number={3},
  pages={1--29},
  year={2024},
  publisher={ACM New York, NY}
}

@incollection{dshalalow2023anthology,
  title={An anthology of classical queueing methods},
  author={Dshalalow, Jewgeni H},
  booktitle={Advances in Queueing Theory, Methods, and Open Problems},
  pages={1--42},
  year={2023},
  publisher={CRC Press}
}

@article{dave2019dmazerunner,
  title={Dmazerunner: Executing perfectly nested loops on dataflow accelerators},
  author={Dave, Shail and Kim, Youngbin and Avancha, Sasikanth and Lee, Kyoungwoo and Shrivastava, Aviral},
  journal={ACM Transactions on Embedded Computing Systems (TECS)},
  volume={18},
  number={5s},
  pages={1--27},
  year={2019},
  publisher={ACM New York, NY, USA}
}

@book{davis2022finite,
  title={Finite State Machine Datapath Design, Optimization, and Implementation},
  author={Davis, Justin and Reese, Robert},
  year={2022},
  publisher={Springer Nature}
}

@INPROCEEDINGS{6691121,
  author={Zhang, Zhiru and Liu, Bin},
  booktitle={2013 IEEE/ACM International Conference on Computer-Aided Design (ICCAD)},
  title={SDC-based modulo scheduling for pipeline synthesis},
  year={2013},
  volume={},
  number={},
  pages={211-218},
  doi={10.1109/ICCAD.2013.6691121}}

@inproceedings{asifuzzaman2021demystifying,
  title={Demystifying the characteristics of high bandwidth memory for real-time systems},
  author={Asifuzzaman, Kazi and Abuelala, Mohamed and Hassan, Mohamed and Cazorla, Francisco J},
  booktitle={2021 IEEE/ACM International Conference On Computer Aided Design (ICCAD)},
  pages={1--9},
  year={2021},
  organization={IEEE}
}

@article{jimenez2024hipporag,
  title={HippoRAG: Neurobiologically Inspired Long-Term Memory for Large Language Models},
  author={Jimenez Gutierrez, Bernal and Shu, Yiheng and Gu, Yu and Yasunaga, Michihiro and Su, Yu},
  journal={Advances in Neural Information Processing Systems},
  volume={37},
  pages={59532--59569},
  year={2024}
}

@inproceedings{stasytis2023optimization,
  title={Optimization Techniques for Hestenes-Jacobi SVD on FPGAs},
  author={Stasytis, Lukas and Istv{\'a}n, Zsolt},
  booktitle={2023 33rd International Conference on Field-Programmable Logic and Applications (FPL)},
  pages={144--150},
  year={2023},
  organization={IEEE}
}

@misc{yu2022orlojpredictablyservingunpredictable,
      title={Orloj: Predictably Serving Unpredictable DNNs},
      author={Peifeng Yu and Yuqing Qiu and Xin Jin and Mosharaf Chowdhury},
      year={2022},
      eprint={2209.00159},
      archivePrefix={arXiv},
      primaryClass={cs.DC},
      url={https://arxiv.org/abs/2209.00159},
}

@misc{dellacroce2018longestprocessingtimerule,
      title={Longest Processing Time rule for identical parallel machines revisited},
      author={Federico Della Croce and Rosario Scatamacchia},
      year={2018},
      eprint={1801.05489},
      archivePrefix={arXiv},
      primaryClass={cs.DS},
      url={https://arxiv.org/abs/1801.05489},
}

@misc{sun2023neardelayoptimalschedulingbatch,
      title={Near Delay-Optimal Scheduling of Batch Jobs in Multi-Server Systems},
      author={Yin Sun and C. Emre Koksal and Ness B. Shroff},
      year={2023},
      eprint={2309.16880},
      archivePrefix={arXiv},
      primaryClass={cs.NI},
      url={https://arxiv.org/abs/2309.16880},
}

@inproceedings{GraphPulse2020,
  author = {S. Rahman and N. Abu-Ghazaleh and R. Gupta},
  title = {GraphPulse: An event-driven hardware accelerator for asynchronous graph processing},
  booktitle = {Proceedings of the 53rd IEEE/ACM International Symposium on Microarchitecture (MICRO-53)},
  year = {2020},
  pages = {908--921}
}

@inproceedings{JetStream2021,
  author = {S. Rahman and M. Afarin and N. Abu-Ghazaleh and R. Gupta},
  title = {JetStream: Graph analytics on streaming data with event-driven hardware accelerator},
  booktitle = {Proceedings of the 54th IEEE/ACM International Symposium on Microarchitecture (MICRO-54)},
  year = {2021},
  pages = {1091--1105}
}

@inproceedings{ScalaGraph2022,
  author = {P. Yao and others},
  title = {ScalaGraph: A scalable accelerator for massively parallel graph processing},
  booktitle = {Proceedings of the 28th IEEE International Symposium on High-Performance Computer Architecture (HPCA)},
  year = {2022},
  pages = {199--212}
}

@INPROCEEDINGS{1203053,
  author={Ying Lu and Abdelzaher, T. and Chenyang Lu and Lui Sha and Xue Liu},
  booktitle={The 9th IEEE Real-Time and Embedded Technology and Applications Symposium, 2003. Proceedings.},
  title={Feedback control with queueing-theoretic prediction for relative delay guarantees in web servers},
  year={2003},
  volume={},
  number={},
  pages={208-217},
  doi={10.1109/RTTAS.2003.1203053}}

@article{choudhury2023accelerating,
  title={Accelerating graph computations on 3D NoC-enabled PIM architectures},
  author={Choudhury, Dwaipayan and Xiang, Lizhi and Rajam, Aravind and Kalyanaraman, Anantharaman and Pande, Partha Pratim},
  journal={ACM Transactions on Design Automation of Electronic Systems},
  volume={28},
  number={3},
  pages={1--16},
  year={2023},
  publisher={ACM New York, NY}
}

@inproceedings{niu2022flashwalker,
  title={FlashWalker: An in-storage accelerator for graph random walks},
  author={Niu, Fuping and Yue, Jianhui and Shen, Jiangqiu and Liao, Xiaofei and Liu, Haikun and Jin, Hai},
  booktitle={2022 IEEE International Parallel and Distributed Processing Symposium (IPDPS)},
  pages={1063--1073},
  year={2022},
  organization={IEEE}
}

@inproceedings{liu2016hierarchical,
  title={Hierarchical random walk inference in knowledge graphs},
  author={Liu, Qiao and Jiang, Liuyi and Han, Minghao and Liu, Yao and Qin, Zhiguang},
  booktitle={Proceedings of the 39th International ACM SIGIR conference on Research and Development in Information Retrieval},
  pages={445--454},
  year={2016}
}

@inproceedings{wang2020learning,
  title={Learning and reasoning on graph for recommendation},
  author={Wang, Xiang and He, Xiangnan and Chua, Tat-Seng},
  booktitle={Proceedings of the 13th international conference on web search and data mining},
  pages={890--893},
  year={2020}
}

@article{10.1145/3588944,
author = {Tan, Hongshi and Chen, Xinyu and Chen, Yao and He, Bingsheng and Wong, Weng-Fai},
title = {LightRW: FPGA Accelerated Graph Dynamic Random Walks},
year = {2023},
issue_date = {May 2023},
publisher = {Association for Computing Machinery},
address = {New York, NY, USA},
volume = {1},
number = {1},
url = {https://doi.org/10.1145/3588944},
doi = {10.1145/3588944},
journal = {Proc. ACM Manag. Data},
month = {may},
articleno = {90},
numpages = {27}
}

@article{liao2023efficient,
  title={Efficient personalized pagerank computation: The power of variance-reduced monte carlo approaches},
  author={Liao, Meihao and Li, Rong-Hua and Dai, Qiangqiang and Chen, Hongyang and Qin, Hongchao and Wang, Guoren},
  journal={Proceedings of the ACM on Management of Data},
  volume={1},
  number={2},
  pages={1--26},
  year={2023},
  publisher={ACM New York, NY, USA}
}

@inproceedings{10.1145/3318464.3380562,
author = {Shao, Yingxia and Huang, Shiyue and Miao, Xupeng and Cui, Bin and Chen, Lei},
title = {Memory-Aware Framework for Efficient Second-Order Random Walk on Large Graphs},
year = {2020},
isbn = {9781450367356},
publisher = {Association for Computing Machinery},
address = {New York, NY, USA},
url = {https://doi.org/10.1145/3318464.3380562},
doi = {10.1145/3318464.3380562},
booktitle = {Proceedings of the 2020 ACM SIGMOD International Conference on Management of Data},
pages = {1797–1812},
numpages = {16},
location = {Portland, OR, USA},
series = {SIGMOD '20}
}

@article{cappelletti2023grape,
  title={GRAPE for fast and scalable graph processing and random-walk-based embedding},
  author={Cappelletti, Luca and Fontana, Tommaso and Casiraghi, Elena and Ravanmehr, Vida and Callahan, Tiffany J and Cano, Carlos and Joachimiak, Marcin P and Mungall, Christopher J and Robinson, Peter N and Reese, Justin and others},
  journal={Nature Computational Science},
  volume={3},
  number={6},
  pages={552--568},
  year={2023},
  publisher={Nature Publishing Group US New York}
}

@inproceedings{10.1145/3543507.3583474,
author = {Li, Zihao and Fu, Dongqi and He, Jingrui},
title = {Everything Evolves in Personalized PageRank},
year = {2023},
isbn = {9781450394161},
publisher = {Association for Computing Machinery},
address = {New York, NY, USA},
url = {https://doi.org/10.1145/3543507.3583474},
doi = {10.1145/3543507.3583474},
booktitle = {Proceedings of the ACM Web Conference 2023},
pages = {3342–3352},
numpages = {11},
location = {, Austin, TX, USA, },
series = {WWW '23}
}

@inproceedings{li2015random,
  title={On random walk based graph sampling},
  author={Li, Rong-Hua and Yu, Jeffrey Xu and Qin, Lu and Mao, Rui and Jin, Tan},
  booktitle={2015 IEEE 31st international conference on data engineering},
  pages={927--938},
  year={2015},
  organization={IEEE}
}

@inproceedings{kunegis2013konect,
  title={Konect: the koblenz network collection},
  author={Kunegis, J{\'e}r{\^o}me},
  booktitle={Proceedings of the 22nd international conference on world wide web},
  pages={1343--1350},
  year={2013}
}

@inproceedings{chen2022regraph,
  title={ReGraph: Scaling graph processing on HBM-enabled FPGAs with heterogeneous pipelines},
  author={Chen, Xinyu and Chen, Yao and Cheng, Feng and Tan, Hongshi and He, Bingsheng and Wong, Weng-Fai},
  booktitle={2022 55th IEEE/ACM International Symposium on Microarchitecture (MICRO)},
  pages={1342--1358},
  year={2022},
  organization={IEEE}
}

@article{walker1974new,
  title={New fast method for generating discrete random numbers with arbitrary frequency distributions},
  author={Walker, Alastair J},
  journal={Electronics Letters},
  volume={8},
  number={10},
  pages={127--128},
  year={1974}
}

@article{zhang2023efficient,
  title={Efficient Dynamic Weighted Set Sampling and Its Extension},
  author={Zhang, Fangyuan and Jiang, Mengxu and Wang, Sibo},
  journal={Proceedings of the VLDB Endowment},
  volume={17},
  number={1},
  pages={15--27},
  year={2023},
  publisher={VLDB Endowment}
}

@article{hou2021massively,
  title={Massively parallel algorithms for personalized pagerank},
  author={Hou, Guanhao and Chen, Xingguang and Wang, Sibo and Wei, Zhewei},
  journal={Proceedings of the VLDB Endowment},
  volume={14},
  number={9},
  pages={1668--1680},
  year={2021},
  publisher={VLDB Endowment}
}

@inproceedings{gao2023fastrw,
  title={FastRW: A Dataflow-Efficient and Memory-Aware Accelerator for Graph Random Walk on FPGAs},
  author={Gao, Yingxue and Wang, Teng and Gong, Lei and Wang, Chao and Li, Xi and Zhou, Xuehai},
  booktitle={2023 Design, Automation \& Test in Europe Conference \& Exhibition (DATE)},
  pages={1--6},
  year={2023},
  organization={IEEE}
}

@article{dgl,
  author       = {Minjie Wang and
                  Lingfan Yu and
                  Da Zheng and
                  Quan Gan and
                  Yu Gai and
                  Zihao Ye and
                  Mufei Li and
                  Jinjing Zhou and
                  Qi Huang and
                  Chao Ma and
                  Ziyue Huang and
                  Qipeng Guo and
                  Hao Zhang and
                  Haibin Lin and
                  Junbo Zhao and
                  Jinyang Li and
                  Alexander J. Smola and
                  Zheng Zhang},
  title        = {Deep Graph Library: Towards Efficient and Scalable Deep Learning on
                  Graphs},
  journal      = {CoRR},
  volume       = {abs/1909.01315},
  year         = {2019},
  url          = {http://arxiv.org/abs/1909.01315},
  eprinttype    = {arXiv},
  eprint       = {1909.01315},
  bibsource    = {dblp computer science bibliography, https://dblp.org}
}

@inproceedings{yin2022scalable,
  title={Scalable Graph Sampling on GPUs with Compressed Graph},
  author={Yin, Hongbo and Shao, Yingxia and Miao, Xupeng and Li, Yawen and Cui, Bin},
  booktitle={Proceedings of the 31st ACM International Conference on Information \& Knowledge Management},
  pages={2383--2392},
  year={2022}
}

@inproceedings{jangda2021accelerating,
  title={Accelerating graph sampling for graph machine learning using GPUs},
  author={Jangda, Abhinav and Polisetty, Sandeep and Guha, Arjun and Serafini, Marco},
  booktitle={Proceedings of the Sixteenth European Conference on Computer Systems},
  pages={311--326},
  year={2021}
}

@inproceedings{gong2023gsampler,
  title={gSampler: General and Efficient GPU-based Graph Sampling for Graph Learning},
  author={Gong, Ping and Liu, Renjie and Mao, Zunyao and Cai, Zhenkun and Yan, Xiao and Li, Cheng and Wang, Minjie and Li, Zhuozhao},
  booktitle={Proceedings of the 29th Symposium on Operating Systems Principles},
  pages={562--578},
  year={2023}
}

@InProceedings{lutz:sigmod:2020,
  author        = {Clemens Lutz and Sebastian Bre{\ss} and Steffen Zeuch and
                  Tilmann Rabl and Volker Markl},
  title         = {Pump up the volume: {Processing} large data on {GPUs} with
                  fast interconnects},
  booktitle     = {{SIGMOD}},
  pages         = {1633--1649},
  publisher     = {{ACM}},
  address       = {New York, NY, USA},
  year          = {2020},
  doi           = {10.1145/3318464.3389705}
}

@inproceedings{chen2021thundergp,
  title={ThunderGP: HLS-based graph processing framework on fpgas},
  author={Chen, Xinyu and Tan, Hongshi and Chen, Yao and He, Bingsheng and Wong, Weng-Fai and Chen, Deming},
  booktitle={The 2021 ACM/SIGDA International Symposium on Field-Programmable Gate Arrays},
  pages={69--80},
  year={2021}
}

@article{chen2022thundergpr,
  title={ThunderGP: Resource-Efficient Graph Processing Framework on FPGAs with HLS},
  author={Chen, Xinyu and Cheng, Feng and Tan, Hongshi and Chen, Yao and He, Bingsheng and Wong, Weng-Fai and Chen, Deming},
  journal={ACM Transactions on Reconfigurable Technology and Systems (TRETS)},
  year={2022},
  publisher={ACM New York, NY}
}

@inproceedings{dong2017metapath2vec,
  title={metapath2vec: Scalable representation learning for heterogeneous networks},
  author={Dong, Yuxiao and Chawla, Nitesh V and Swami, Ananthram},
  booktitle={Proceedings of the 23rd ACM SIGKDD international conference on knowledge discovery and data mining},
  pages={135--144},
  year={2017}
}

@article{sun2021thunderrw,
  title={ThunderRW: an in-memory graph random walk engine},
  author={Sun, Shixuan and Chen, Yuhang and Lu, Shengliang and He, Bingsheng and Li, Yuchen},
  journal={Proceedings of the VLDB Endowment},
  volume={14},
  number={11},
  pages={1992--2005},
  year={2021},
  publisher={VLDB Endowment}
}

@inproceedings{su2021graph,
  title={Graph Sampling with Fast Random Walker on HBM-enabled FPGA Accelerators},
  author={Su, Chunyou and Liang, Hao and Zhang, Wei and Zhao, Kun and Ai, Baole and Shen, Wenting and Wang, Zeke},
  booktitle={2021 31st International Conference on Field-Programmable Logic and Applications (FPL)},
  pages={211--218},
  year={2021},
  organization={IEEE}
}

@inproceedings{grover2016node2vec,
  title={node2vec: Scalable feature learning for networks},
  author={Grover, Aditya and Leskovec, Jure},
  booktitle={Proceedings of the 22nd ACM SIGKDD international conference on Knowledge discovery and data mining},
  pages={855--864},
  year={2016}
}

@inproceedings{yang2019knightking,
  title={Knightking: a fast distributed graph random walk engine},
  author={Yang, Ke and Zhang, MingXing and Chen, Kang and Ma, Xiaosong and Bai, Yang and Jiang, Yong},
  booktitle={Proceedings of the 27th ACM Symposium on Operating Systems Principles},
  pages={524--537},
  year={2019}
}

@misc{snapnets,
  author       = {Jure Leskovec and Andrej Krevl},
  title        = {{SNAP Datasets}: {Stanford} Large Network Dataset Collection},
  howpublished = {\url{http://snap.stanford.edu/data}},
  month        = jun,
  year         = 2014
}

@inbook{thundering,
author = {Tan, Hongshi and Chen, Xinyu and Chen, Yao and He, Bingsheng and Wong, Weng-Fai},
title = {ThundeRiNG: Generating Multiple Independent Random Number Sequences on FPGAs},
year = {2021},
isbn = {9781450383356},
publisher = {Association for Computing Machinery},
address = {New York, NY, USA},
url = {https://doi.org/10.1145/3447818.3461664},
booktitle = {Proceedings of the ACM International Conference on Supercomputing},
pages = {115–126},
numpages = {12}
}

@inproceedings{BoVWFI,
  author = "Paolo Boldi and Sebastiano Vigna",
  title = "The {W}eb{G}raph Framework {I}: {C}ompression Techniques",
  year = 2004,
  booktitle = "Proc. of the Thirteenth International World Wide Web Conference (WWW 2004)",
  address = "Manhattan, USA",
  pages = "595--601",
  publisher = "ACM Press"
}

@article{zhou2019hitgraph,
  title={Hit{G}raph: High-throughput graph processing framework on {FPGA}},
  author={Zhou, Shijie and Kannan, Rajgopal and Prasanna, Viktor K and Seetharaman, Guna and Wu, Qing},
  journal={IEEE Transactions on Parallel and Distributed Systems},
  volume={30},
  number={10},
  pages={2249--2264},
  year={2019},
  publisher={IEEE}

}

@inproceedings{moreira2020neuronflow,
  title={NeuronFlow: A Hybrid Neuromorphic--Dataflow Processor Architecture for AI Workloads},
  author={Moreira, Orlando and Yousefzadeh, Amirreza and Chersi, Fabian and Kapoor, Ajay and Zwartenkot, R-J and Qiao, Peng and Cinserin, Gokturk and Khoei, Mina A and Lindwer, Menno and Tapson, Jonathan},
  booktitle={2020 2nd IEEE International Conference on Artificial Intelligence Circuits and Systems (AICAS)},
  pages={01--05},
  year={2020},
  organization={IEEE}
}

@inproceedings{miller2013graph,
  title={Graph database applications and concepts with Neo4j},
  author={Miller, Justin J},
  booktitle={Proceedings of the southern association for information systems conference, Atlanta, GA, USA},
  volume={2324},
  pages={141--147},
  year={2013}
}

@inproceedings{chakrabarti2004r,
  title={R-MAT: A recursive model for graph mining},
  author={Chakrabarti, Deepayan and Zhan, Yiping and Faloutsos, Christos},
  booktitle={Proceedings of the 2004 SIAM International Conference on Data Mining},
  pages={442--446},
  year={2004},
  organization={SIAM}
}

@inproceedings{hu2021graphlily,
  title={GraphLily: Accelerating graph linear algebra on HBM-equipped FPGAs},
  author={Hu, Yuwei and Du, Yixiao and Ustun, Ecenur and Zhang, Zhiru},
  booktitle={2021 IEEE/ACM International Conference On Computer Aided Design (ICCAD)},
  pages={1--9},
  year={2021},
  organization={IEEE}
}

@inproceedings{perozzi2014deepwalk,
  title={Deepwalk: Online learning of social representations},
  author={Perozzi, Bryan and Al-Rfou, Rami and Skiena, Steven},
  booktitle={Proceedings of the 20th ACM SIGKDD international conference on Knowledge discovery and data mining},
  pages={701--710},
  year={2014}
}

@inproceedings{yu2014personalized,
  title={Personalized entity recommendation: A heterogeneous information network approach},
  author={Yu, Xiao and Ren, Xiang and Sun, Yizhou and Gu, Quanquan and Sturt, Bradley and Khandelwal, Urvashi and Norick, Brandon and Han, Jiawei},
  booktitle={Proceedings of the 7th ACM international conference on Web search and data mining},
  pages={283--292},
  year={2014}
}

@misc{hacc,
  author  = {AMD},
  title   = {Heterogeneous Accelerated Compute Clusters (HACC) Program},
  howpublished     = {\url{https://www.amd-haccs.io/index.html}},
  year = {2023},
}

\end{document}